\documentclass[11pt]{article}

\usepackage{graphicx}
\usepackage{cite}
\usepackage{amssymb}
\usepackage{amsmath}
\usepackage[colorlinks=true]{hyperref}
\def\beq{\begin{equation}}
\def\eeq{\end{equation}}
\def\ba{\begin{array}}
\def\ea{\end{array}}
\def\bea{\begin{eqnarray}}
\def\eea{\end{eqnarray}}

\def\sq2{\sqrt{2}}
\def\sqs{\sqrt{s}}
\def\ab1{{\rm ab}^{-1}}

\newcommand{\gm}{\gamma^\mu}
\newcommand{\smn}{\sigma^{\mu \nu}}

\newcommand{\vphi}{\varphi}

\newcommand{\darrd}{\stackrel{\leftrightarrow}{D}}
\newcommand{\darrdk}{\stackrel{\leftrightarrow}{D^k}}

\newcommand{\RI}{\ensuremath{R_{<2.5}}}
\newcommand{\RII}{\ensuremath{R_{<4.0}}}
\newcommand{\RIII}{\ensuremath{R_{<6.0}}}
\newcommand{\RIc}{\ensuremath{R_{>2.5}}}

\newcommand{\toptag}{\ensuremath{\operatorname{top-tag}}}
\newcommand{\higgstag}{\ensuremath{\operatorname{higgs-tag}}}

\begin{document}
\author{Nicol\'as Urcelay\thanks{nicolas.urcelay@cinvestav.mx}, \
  Antonio O. Bouzas\thanks{abouzas@cinvestav.mx} \
  and F. Larios\thanks{francisco.larios@cinvestav.mx,
  corresponding author.}
\\
Departamento de F\'{\i}sica Aplicada,
CINVESTAV-M\'erida, \\ A.P. 73, 97310 M\'erida,
Yucat\'an, M\'exico}

\title{Top quark couplings at the muon collider}

\date{\today}


\maketitle

\begin{abstract}

We present an overview of the sensitivity of a muon collider to the
flavor-diagonal electroweak couplings of the top quark, using
dimension-six operators of the Warsaw basis.  Of the nine
fermion-boson operators involving the top quark, the bounds on
three are not competitive with existing limits, and we focus on the
remaining six.  We estimate individual $95\%$ CL limits
on the corresponding Wilson coefficients at $\sqs = 3$, $6$, $10$,
and $14$~TeV from $2\sigma$ statistical cross-section uncertainties,
and assess the effect of an assumed systematic uncertainty.  We
consider six top-quark production modes: 
  $t\bar t$, $t \bar t h$,
  $t \bar t \nu \bar \nu$, $t \bar t \mu^+ \mu^-$, single-top
  $t\bar b \mu^- \bar \nu_\mu$, and single-top-plus-Higgs, the last
  three for several rapidity coverages of the final-state muons.  The
  $s$-channel $t\bar t$ and $t\bar t h$ processes give the strongest
  sensitivity to the electroweak dipole couplings, with allowed
  intervals narrower than the current LHC ones by factors of about $3$
  to $6$ at $\sqs = 3$ and $6$~TeV, and by an order of magnitude or
  more at $10$ and $14$~TeV.  The left- and right-handed $ttZ$
  couplings, as well as the Yukawa $tth$ couplings (for
  $\sqs \geq 6$~TeV), are best probed in $t \bar t \nu \bar \nu$
  production, while the right-handed $tbW$ coupling is best probed in
  single-top production.  For the $ttZ$ and $tbW$ couplings, the
  limits improve on the current LHC bounds by factors from about $1.5$
  to more than an order of magnitude.  Our results agree with previous
  analyses of the Yukawa and the dipole couplings, and provide a
  roadmap for more in-depth studies.
\end{abstract}


\section{Introduction}

The Large Hadron Collider (LHC) program has probed Higgs-boson and
top-quark interactions, and has found all observables to be
compatible with the Standard Model (SM).  The high-luminosity phase
of the LHC (HL-LHC) is expected to improve these analyses somewhat,
thanks to the increase in data.  However, there is a recognized need
to probe electroweak interactions at high energies in the cleaner
environment of a lepton collider.  Much attention has been given to
the Muon Collider (MuC), since such a machine would combine the
energy reach of a hadron collider with the precision of a lepton
collider~\cite{muonsmasher,muc22,mucreport22,muc23}.

We are interested in the sensitivity that the MuC can reach at
different energies for the couplings of the top quark to gauge and
Higgs bosons generated by the dimension-six operators of the Warsaw
basis~\cite{grz10}.  We consider the flavor-diagonal operators only;
for flavor-changing top-quark couplings at the MuC, we refer to
Refs.~\cite{ake23,moretti26,senol26,airen26,glioti25}.
We also do not consider four-fermion operators, which have been
addressed in Ref.~\cite{mele26}.
A previous study, Ref.~\cite{han24}, has addressed two of the
couplings (the electroweak dipole operators), and yet another,
Ref.~\cite{miralles2503}, has given some limits on three others.
Here, we aim for a comprehensive analysis of the nine possible Wilson
coefficients of interest and of all the most relevant top-quark
production processes.  For three of them, $C^{(+)}_{\vphi Q}$,
$C_{tG}$, and $C_{bW}$, top production at the MuC is not competitive
with existing limits (see Sect.~\ref{sec:operators}), and we focus on
the remaining six.

Currently, two center-of-mass energies in particular are considered
desirable and realistic for the MuC project: $\sqs = 3$ and
$10$~TeV~\cite{MuC25}.  In this study, we also include
$\sqs = 6$ and $14$~TeV, to obtain a better view of the energy
dependence.  For the integrated luminosity, we use the well-known
formula ${\cal L} = E^2_{\rm cm}/10$, with ${\cal L}$ in units of
$\ab1$ and $E_{\rm cm}$ in TeV~\cite{muc22}.  In
Table~\ref{tab:lumin}, we show the four energies and the
corresponding integrated luminosities.

\begin{table}[ht!]
  \centering
  \begin{tabular}{|r|c|c|c|c|}\hline
    $\sqs$ \ [TeV]  & $3$ & $6$ & $10$ & $14$
  \\\hline
  ${\cal L} \ [\ab1]$ & $0.9$ & $3.6$ & $10$ & $19.6$
\\\hline
  \end{tabular}                               
  \caption{Benchmark values for energy and luminosity of a
    high energy MuC used in this study. Luminosity given by
    the equation $(1)$ in \cite{muc22}.}
  \label{tab:lumin}
\end{table}

Higgs physics and the nature of electroweak symmetry breaking (EWSB)
are the main focus of interest in the literature, especially at $\sqs
= 10$ TeV where $WW$ and $ZZ$ fusion production modes of the Higgs 
boson become dominant. The second mode involves outgoing muons with
rapidities in the range $3 \leq |y| < 6$, which makes the case
for tagging forward muons~\cite{meade22}.  For a recent and thorough
study of the opportunities offered by tagging forward muons, we
refer to Ref.~\cite{ruhdorfer24} (see also
Refs.~\cite{ruhdorfer19,lily24,lima20}).
The European Strategy for Particle Physics (2026 update) is
already considering such a capability as \emph{aspirational}, and it
focuses on detecting muons with rapidities up to $|y| = 4$ or
$|y| = 6$~\cite{MuC25}.  In this study, we consider final-state
muons in the ranges $|y| < 2.5$, $|y| < 4$, and
$|y| < 6$ for processes that involve neutral-gauge-boson fusion
diagrams.

The article is organized as follows.  In Sect.~\ref{sec:operators},
we present the dimension-six fermion-boson operators involving the top
quark that are relevant for this study.  In
Sect.~\ref{sec:monte.carlo}, we discuss the most important
tree-level production modes with top quarks in the final state:
$\mu^+ \mu^- \to t \bar t$, $t \bar t h$,
$t \bar t \nu \bar \nu$, $t \bar t \mu^+ \mu^-$,
$t \bar b \mu^- \bar \nu_\mu $, $t \bar b \mu^- \bar \nu_\mu h$.  Based
on $2\sigma$ cross-section uncertainties, we estimate $95\%$
confidence-level (CL) limits in Sect.~\ref{sec:limits}, where we
identify the processes with the highest sensitivity for each of the
coefficients and, in subsection~\ref{sec:marginalized}, provide
two-dimensional allowed regions and marginalized limits for the
electroweak dipole coefficients.  In Sect.~\ref{sec:conclusions}, we
present a summary and our final remarks.
Appendix~\ref{app:limits} gives the individual limits on the six
coefficients of interest, for both the real and imaginary parts of
the complex ones; in each case, we show the limits set by the five
most sensitive processes for that coefficient, together with the
effect of an assumed systematic uncertainty, which can be significant
for processes with large cross sections.


\section{Dimension-six fermion-boson operators involving the top quark}
\label{sec:operators}
We refer to the list of dimension-six effective operators
known as the Warsaw basis in Ref.~\cite{grz10} and to a subsequent
article~\cite{feynrules}, where the associated Feynman rules
(in $R_\xi$ gauges) are provided.  Two types of
operators involving the top quark are recognized: fermion-boson
and four-fermion.  We will not consider the latter; for a recent
study of the sensitivity to four-fermion interactions involving two
leptons and two top quarks at the MuC, we refer to
Ref.~\cite{mele26}.
There are nine fermion-boson operators where up-type quarks appear:
\begin{eqnarray}
Q_{\vphi u} &=& \vphi^\dagger i {\darrd}_\mu \vphi \;
\bar u \gm u \; ,\nonumber \\
Q^{(1)}_{\vphi q} &=& (\vphi^\dagger i {\darrd}_\mu \vphi)
\bar q \gm q  \; ,\nonumber \\
Q^{(3)}_{\vphi q} &=& (\vphi^\dagger i {\darrdk}_\mu \vphi)
\bar q \tau^k \gm q  \; ,\nonumber \\
Q_{\vphi ud} &=& ({\tilde \vphi}^\dagger i {\darrd}_\mu \vphi)
\bar u \gm d 
\; ,\label{eq:operators} \\
Q_{u\vphi} &=& \vphi^\dagger \vphi \; \bar q  u {\tilde \vphi}
\; ,\nonumber \\
Q_{uB} &=& \bar q \smn u {\tilde \vphi} B_{\mu \nu} \; ,
\nonumber \\
Q_{uW} &=& \bar q \smn u \tau^k {\tilde \vphi} W^k_{\mu \nu} 
\; ,\nonumber \\
Q_{dW} &=& \bar q \smn d \tau^k {\vphi} W^k_{\mu \nu} 
\; ,\nonumber \\
Q_{uG} &=& \bar q \smn u T^A {\tilde \vphi} G^A_{\mu \nu} 
\; .\nonumber 
\end{eqnarray}
We omit family indices, as we are only interested in flavor-diagonal 
couplings of the third family. As is common practice in the 
literature, we will refer to the right-handed up-type quark $u$
of the third family as $t$, 
and use $q$ for the left-handed doublets of the first two generations 
and $Q$ for that of the third. Therefore, we will write $Q_{\vphi u}$ 
as $Q_{\vphi t}$, $Q_{\vphi ud}$ as $Q_{\vphi tb}$, 
$Q^{(1)}_{\vphi q}$ as $Q^{(1)}_{\vphi Q}$, and so on. The Lagrangian 
we use is written as:
\begin{equation}
  \begin{aligned}
    {\cal L} &= {\cal L}_{\rm SM}  + \frac{1}{\Lambda^2} \left[
      {C}_{\vphi t} Q_{\vphi t} + {C}^{(1)}_{\vphi Q} Q^{(1)}_{\vphi Q}
      + {C}^{(3)}_{\vphi Q} Q^{(3)}_{\vphi Q} \right] \\
    & \quad + \frac{1}{\Lambda^2} \left[ {C}_{\vphi tb} Q_{\vphi tb} +
      {C}_{t\vphi} Q_{t\vphi} + {C}_{tB} Q_{tB} + {C}_{tW} Q_{tW} +
      {C}_{bW} Q_{bW} + {C}_{tG} Q_{tG} + {\rm H.c.}
      \rule{0pt}{11pt}\right].    
  \end{aligned}
  \label{eq:smeft}
\end{equation}
Notice that the first three operators in (\ref{eq:operators}) are 
Hermitian; correspondingly, the Wilson coefficients on the first 
line of (\ref{eq:smeft}) are real, whereas those on the second line 
are complex. We set the mass scale $\Lambda = 1$~TeV for convenience.

Concerning the operators $Q^{(1)}_{\vphi Q}$ and $Q^{(3)}_{\vphi Q}$,
the following combinations are in common use in 
the literature~\cite{zhang18, aguilargroup18}:
\begin{equation}
  Q^{(\pm)}_{\vphi Q} \equiv \frac{1}{2} \left[ Q^{(1)}_{\vphi Q} \pm
    Q^{(3)}_{\vphi Q} \right],
  \quad
  C^{(\pm)}_{\vphi Q} \equiv C^{(1)}_{\vphi Q} \pm
  C^{(3)}_{\vphi Q},
\end{equation}
so that
\begin{equation}
  {C}^{(1)}_{\vphi Q} Q^{(1)}_{\vphi Q}
  + {C}^{(3)}_{\vphi Q} Q^{(3)}_{\vphi Q} =
  {C}^{(-)}_{\vphi Q} Q^{(-)}_{\vphi Q}
  + {C}^{(+)}_{\vphi Q} Q^{(+)}_{\vphi Q}.
\end{equation}
Both $Q^{(+)}_{\vphi Q}$ and $Q^{(-)}_{\vphi Q}$ generate a
left-handed $tbW$ coupling, but
$Q^{(-)}_{\vphi Q}$ generates a $ttZ$ vertex, 
whereas $Q^{(+)}_{\vphi Q}$ generates a $Zbb$ coupling.
It follows that $C^{(+)}_{\vphi Q}$ is strongly bounded at the tree 
level by electroweak precision data from LEP and SLC. 
A recent limit can be found in Ref.~\cite{zhang22}: 
$-0.015 \leq C^{(+)}_{\vphi Q} \leq +0.019$ (Table 6, $68\%$ CL). 
It is important to bear this separation in mind: neglecting it can
lead to artificially strong individual constraints on 
$C^{(1)}_{\vphi Q}$ and $C^{(3)}_{\vphi Q}$ when they are obtained
from global fits that include LEP and SLC measurements. 
For example, let us consider $C^{(3)}_{\vphi Q}$. In a 2025 publication, 
the ATLAS collaboration reports individual limits of $[-0.45, 0.13]$ 
($68\%$ CL) based on top-quark processes recorded by the ATLAS 
detector (see Table~7 in Ref.~\cite{atlas25}). 
In a 2026 publication, they report a much smaller bound of 
$\pm 0.02$ ($68\%$ CL), which is due to the inclusion of LEP 
and SLC data (see Fig.~7 in Ref.~\cite{atlas26}). 
A similar situation can be found in 
Refs.~\cite{cms25, miralles2507, miralles2512, miralles2503}. 
It should be noted that $C^{(-)}_{\vphi Q}$ can also be bounded by 
electroweak data, but the sensitivity comes from contributions 
at the one-loop level. In this article, we are interested in 
the sensitivity from direct, tree-level contributions.

\begin{table} \centering{}
  \begin{tabular}{|c|c|c|c|c|}
    \cline{1-5}
    $C^{}_{bW}$ \cite{cms23} & $C^{}_{tG}$ \cite{cms23} &
  $C^{}_{tB}$ \cite{cms25} & $C^{}_{tW}$ \cite{cms25} &
    $C^{}_{t\vphi}$ \cite{atlas26} \\
    $[-0.75,0.75]$ & $[-0.21,0.24]$ &
    $[-0.15,0.016]$ &$[-0.26,0.034]$ &
    $[-1.6,2.8]$\\
\hline
$C^{}_{\vphi tb}$ \cite{cms23} & $C^{}_{\vphi t}$ \cite{atlas23} &
$C^{(1)}_{\vphi Q}$ \cite{atlas23} & $C^{(3)}_{\vphi Q}$\cite{cms23} & 
$C^{(-)}_{\vphi Q}$ \cite{cms23}\\
$[-3.14,3.18]$ & $[-2.20,1.60]$ & $[-1.40,0.84]$ &
$[-0.85,1.89]$ & $[-2.68,2.94]$\\
    \hline   
  \end{tabular}
  \caption{Individual $95\%$ CL limits on the effective couplings
   from top-quark processes at CMS \cite{cms23,cms25} and
    ATLAS \cite{atlas23,atlas26}.}
  \label{tab:cmsatlaslim}
  \end{table}

There have been a substantial number of articles presenting
limits from the LHC in the last 15 years.
In Table~\ref{tab:cmsatlaslim} we show $95\%$ CL bounds based on
top-quark production as reported by CMS and ATLAS publications
in the last three years.
Since the maximum size of $C^{(+)}_{\vphi Q}$ is about two orders of
magnitude smaller than the limits on $C^{(-)}_{\vphi Q}$, we can
use the approximation $C^{(1)}_{\vphi Q} \simeq -C^{(3)}_{\vphi Q}$,
so that $C^{(-)}_{\vphi Q} \simeq 2 C^{(1)}_{\vphi Q}$.  Indeed, this
relation is consistent with the limits shown in
Table~\ref{tab:cmsatlaslim}.

With the projected $3 \ab1$ of data from the high-luminosity
phase of the LHC (HL-LHC), these limits should be reduced,
but the estimates in the literature vary significantly.
For instance, from Tables~52 and~53 of the 2019 CERN Yellow Report,
Ref.~\cite{hllhcreport1}, we observe estimated $95\%$ CL individual
limits for some of the Wilson coefficients that are actually similar
to the ones already established in Table~\ref{tab:cmsatlaslim}.
In the last two years, significantly more stringent estimates have
been published in Refs.~\cite{miralles2503,smefit26}.
We show them in Table~\ref{tab:hllhc}.  These references use the
combination $C_{tZ} = c_w C_{tW} - s_w C_{tB}$, with
$c_w = \cos\theta_W$, in place of $C_{tB}$.  Similar recent estimates
can also be found in Ref.~\cite{hoeve25}.

Comparing with the LHC bounds, we notice that the dipole couplings
could get remarkably stringent limits of order $10^{-2}$ at the
HL-LHC.  In contrast, we observe that the limits on the Yukawa
coefficient $C_{t \vphi}$ could be reduced by somewhat less than an
order of magnitude.  Table~\ref{tab:hllhc} also shows the marginalized
limits for the same Wilson coefficients presented by
Ref.~\cite{smefit26} in Fig.~68.  We notice a difference of one or two
orders of magnitude, suggesting that the stringency of individual
limits can be misleading. The MuC at $10$~TeV could actually yield
bounds on $C_{tB}$ and $C_{tW}$ (comparable to those on $C_{tZ}$ and
$C_{tW}$) even smaller than the projected individual limits of the
HL-LHC, as we shall see below, but even if they were of similar size,
the input from this collider would provide an important contribution
to the overall fit.

As noted in the Introduction, top production at the MuC is not
competitive for three of the coefficients: $C^{(+)}_{\vphi Q}$,
which is strongly constrained by LEP and SLC data, $C_{tG}$, and
$C_{bW}$.  For the chromomagnetic
coupling $C_{tG}$, CMS \cite{cms23} obtained a bound of
about $|C_{tG}| < 0.2$ in 2023, whereas a global fit performed one
year later in Ref.~\cite{celada24} obtained a significantly smaller
bound of $0.08$.  Furthermore, Ref.~\cite{kidonakis26}, a recent study
of LHC $t\bar t$ production at next-to-next-to-leading order (NNLO),
reports an even smaller individual bound of about $0.05$.  At the MuC,
$t\bar t g$ production could serve as a probe.  We have tested this
option and obtained limits of roughly $0.8$ at all four energies,
about one order of magnitude weaker than the recent LHC bounds.  We
therefore do not address the $C_{tG}$ coupling in the remainder of
this article.
For $C_{bW}$, we obtain limits from single-top production
(Sect.~\ref{sec:limsinglet}), but a recent study
\cite{barducci} reports much stronger limits from $b\bar b$ and
$b\bar b h$ production.

\begin{table} \centering{}
  \begin{tabular}{|c||c|c|c|c|c|}
    \cline{1-6}
  {\cite{hllhcreport1}} & $C^{}_{tB}$ & $C^{}_{tW}$ &
  $C^{}_{tG}$ & $C^{}_{\vphi t}$ & $C^{(1)}_{\vphi Q}$ \\
  {2019} & $[-0.5,0.3]$ & $[-0.3,0.5]$ &
    $[-0.1,0.1]$ &$[-0.89,0.89]$ &
    $[-0.75,0.73]$\\
  \hline
  {\cite{miralles2503}} & $C^{}_{tZ}$ & $C^{}_{tW}$ &
  $C^{}_{tG}$ & $C^{}_{\vphi t}$ & $C^{}_{t \vphi}$ \\
 {2025} & $0.70$ & $0.11$ &
 $0.08$ & $0.60$ & $1.4$\\
 \hline
  {\cite{smefit26}} & $C^{}_{tZ}$ & $C^{}_{tW}$ &
  $C^{}_{tG}$ & $C^{}_{\vphi t}$ & $C^{}_{t \vphi}$ \\
 {2026} & $0.008$ & $0.007$ &
 $0.024$ & $0.075$ & $0.45$\\
 \hline
  {\cite{smefit26}} & $C^{}_{tZ}$ & $C^{}_{tW}$ &
  $C^{}_{tG}$ & $C^{}_{\vphi t}$ & $C^{}_{t \vphi}$ \\
 {2026-m} & $0.28$ & $0.20$ &
    $0.15$ & $2.6$ & $2.0$\\
\hline
  \end{tabular}
  \caption{Three projected $95\%$ CL individual limits on the
    effective couplings at the HL-LHC. The last row shows
  marginalized limits.}
  \label{tab:hllhc}
  \end{table}


\section{Top production processes: Monte Carlo simulations}
\label{sec:monte.carlo}

\begin{figure}[t]
  \centering
  \includegraphics[scale=1.0]{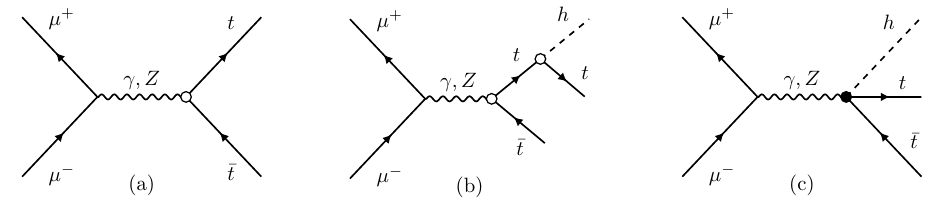}
  \caption{Illustrative unitary-gauge Feynman diagrams for (a)
    top-pair production, (b), (c) associated top-pair plus Higgs
    production. Three types of vertices are distinguished: standard
    unmarked vertices represent SM couplings, solid-circle vertices
    indicate anomalous couplings, open-circle vertices can be either
    SM or anomalous.}
  \label{fig:dgrms.1}
\end{figure}

\begin{figure}[t]
  \centering
  \includegraphics[scale=1.0]{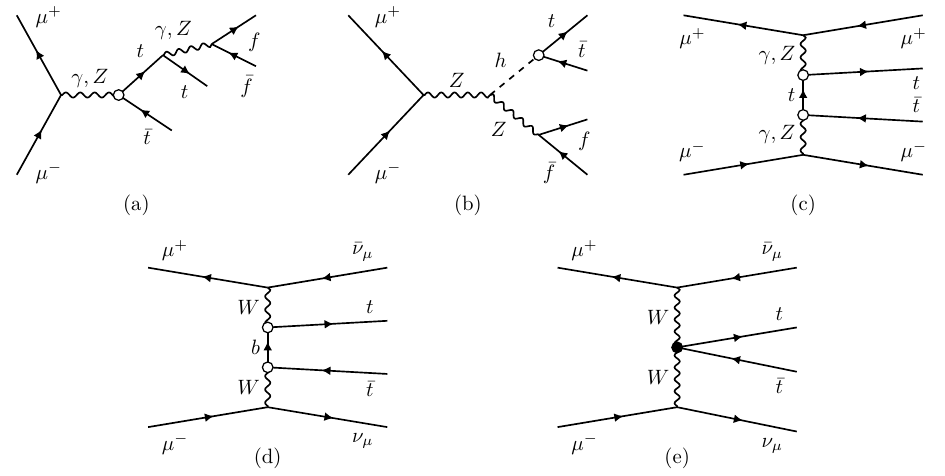}
\caption{Illustrative unitary-gauge Feynman diagrams for
  $\mu^+\mu^-\rightarrow t\bar{t}\mu^+\mu^-$: (a), (b), (c) with
  $f=\mu^-$; $\mu^+\mu^-\rightarrow t\bar{t}\nu\bar{\nu}$: (a), (b),
  (d), (e) with $f=\nu_e,$ $\nu_\mu,$ $\nu_\tau$. Graphical code
  for the vertices as in Fig.~\ref{fig:dgrms.1}.}
  \label{fig:dgrms.2}
\end{figure}

\begin{figure}[t]
  \centering
  \includegraphics[scale=1.0]{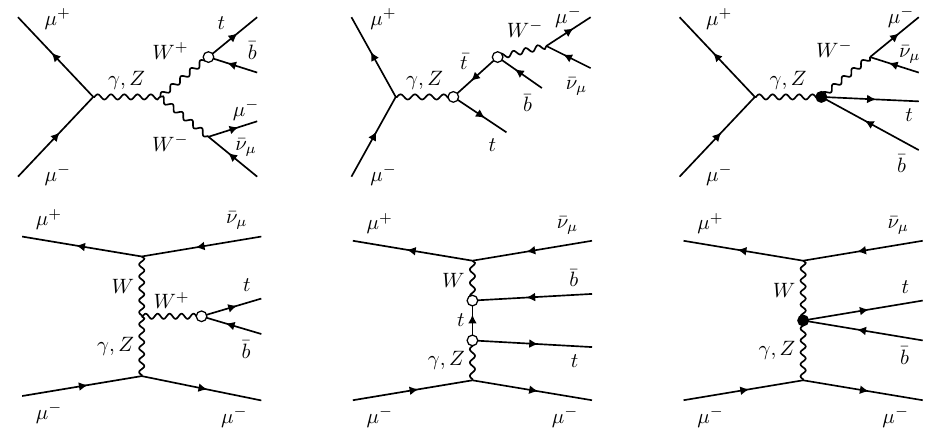}
\caption{Illustrative unitary-gauge Feynman diagrams for single-top
  production $\mu^+\mu^-\rightarrow t\bar{b}\mu^-\bar\nu_\mu$. Upper row:
  annihilation diagrams, lower row: vector boson fusion
  diagrams. Graphical code for the vertices as in
  Fig.~\ref{fig:dgrms.1}.}
  \label{fig:dgrms.3}
\end{figure}

\begin{figure}[t]
  \centering
  \includegraphics[scale=1.0]{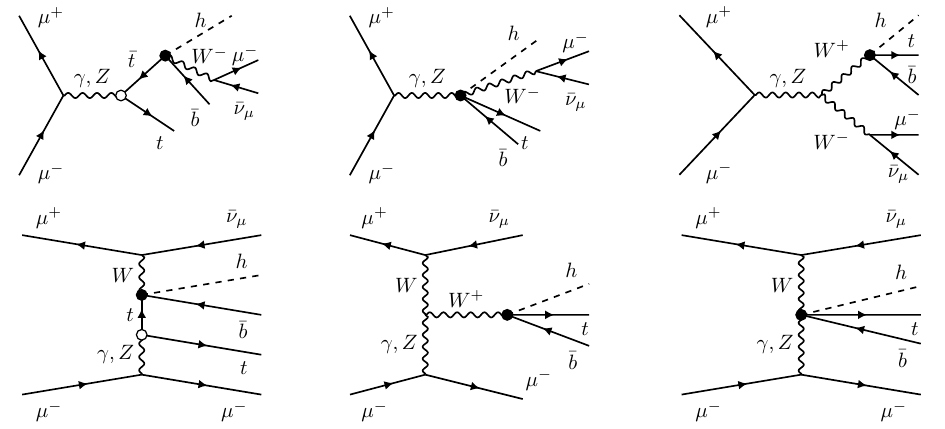}
\caption{Illustrative unitary-gauge Feynman diagrams for associated
  single-top plus Higgs production
  $\mu^+\mu^-\rightarrow t\bar{b}\mu^-\bar\nu_\mu h$. Upper row:
  annihilation diagrams, lower row: vector boson fusion
  diagrams. Diagrams that can be obtained by inserting a Higgs line
  into a massive propagator line via a trivalent vertex in
  Fig.~\ref{fig:dgrms.3} are omitted.  Graphical code for the vertices
  as in Fig.~\ref{fig:dgrms.1}.}
  \label{fig:dgrms.4}
\end{figure}

We consider six top-quark production processes in this work:
\begin{subequations}
  \label{eq:six.procs}  
\begin{align}
\mu^+\mu^- &\rightarrow t\bar{t}, \label{eq:top.pair}\\
\mu^+\mu^- &\rightarrow t\bar{t} h, \label{eq:top.pair.higgs}\\
\mu^+\mu^- &\rightarrow t\bar{t}\nu\bar{\nu}, \label{eq:top.pair.nunu}\\  
\mu^+\mu^- &\rightarrow t\bar{t}\mu^+\mu^-, \label{eq:top.pair.mumu}\\
\mu^+\mu^- &\rightarrow t\bar{b}\mu^-\bar{\nu}_\mu +
\bar{t}b\mu^+\nu_\mu, \label{eq:sngl.top}\\
\mu^+\mu^- &\rightarrow t\bar{b}\mu^-\bar{\nu}_\mu h +
\bar{t}b\mu^+\nu_\mu h. \label{eq:sngl.top.h}  
\end{align}
\end{subequations}
Illustrative unitary-gauge Feynman diagrams for the processes in
Eq.~(\ref{eq:six.procs}) are provided in
Figs.~\ref{fig:dgrms.1}--\ref{fig:dgrms.4}. In
Fig.~\ref{fig:dgrms.1}(a), we show the Feynman diagrams for top-pair
production, which involve the $s$-channel exchange of $\gamma$ and $Z$
bosons that couple to the top quark through standard model (SM) or
dimension-six SMEFT couplings. Also shown in this figure are diagrams
for associated top-pair production with a Higgs boson. In diagram~(b),
the Higgs boson is inserted onto the top-quark line via a trivalent
vertex, which can originate from either the SM or the SMEFT; however,
diagram~(c) involves a tetravalent vertex that does not exist in the
SM.  Figure~\ref{fig:dgrms.2} displays the Feynman diagrams for
$\mu^+\mu^- \rightarrow t \bar{t} \mu^+\mu^-$ [diagrams~(a), (b), and
(c) with $f=\mu^-$] and for
$\mu^+\mu^- \rightarrow t \bar{t} \nu \bar{\nu}$ [diagrams~(a), (b),
(d), and (e) with $f=\nu_e, \nu_\mu, \text{ or } \nu_\tau$].  Only
annihilation diagrams such as (a) and (b) can feature $\nu_e$ or
$\nu_\tau$ in the final state, and their contribution to the SM cross
section is negligibly small: less than 0.2\% at $\sqrt{s}=3$ TeV and
even smaller at higher energies. Vector-boson fusion (VBF) diagrams in
Figs.~\ref{fig:dgrms.2}(d) and \ref{fig:dgrms.2}(e) can only have
$\nu_\mu$ in the final state; furthermore, diagram~(e) contains a
tetravalent vertex that does not exist in the SM.

In Fig.~\ref{fig:dgrms.3}, we show the Feynman diagrams for the
single-top production process
\(\mu^+\mu^- \rightarrow t \bar{b} \mu^- \bar{\nu}_\mu\). Both
annihilation and VBF diagrams are depicted, with the former dominating
the cross section for \(\sqrt{s} \lesssim 2\)~TeV and the latter for
\(\sqrt{s} \gtrsim 2\)~TeV. This crossover reflects the increasingly
important role of electroweak radiation at high energies and is
particularly relevant for SMEFT studies, as annihilation is sensitive
to four-fermion contact operators, whereas VBF probes Higgs-gauge
couplings to fermions. Feynman diagrams for the associated production
of a single top quark and a Higgs boson can be obtained from the
diagrams in Fig.~\ref{fig:dgrms.3} by attaching a Higgs line to any
massive internal line via a trivalent vertex, corresponding to either
an SM or a SMEFT coupling.  In Fig.~\ref{fig:dgrms.4}, we illustrate
diagrams that cannot be obtained in this manner, as they contain a
tetravalent or pentavalent SMEFT vertex that does not exist in the SM.

We simulate the processes in Eq.~(\ref{eq:six.procs}) using
\textsc{MadGraph5\_aMC@NLO} (henceforth MG5) version
3.6.7~\cite{alw14}. We implemented the dimension-six SMEFT operators
from Sect.~\ref{sec:operators} in MG5 by means of \textsc{FeynRules}
version 2.3.49~\cite{all14}.  Furthermore, we analyzed the events
generated by MG5 using \textsc{ROOT} version 6.39.01~\cite{root}.  We
also employed the program \textsc{CalcHEP} version
3.9.2~\cite{calchep} with manually coded anomalous vertices to
cross-check our results. For the numerical calculations in this work,
we adopt $\alpha^{-1}=127.9$, $M_Z=91.1876$~GeV,
$G_F=1.16637\times10^{-5}$~GeV$^{-2}$ (which implies
$\sin^2\theta_W=0.2312$), $m_t=172.5$~GeV, $m_b=4.5$~GeV,
$m_u=m_d=m_s=m_c=m_\mu=m_e=0$, $\theta_C=0.22736$, and
$\Lambda=1$~TeV.

In our simulations, we always apply a rapidity cut on the final-state
particles of the processes in Eq.~(\ref{eq:six.procs}):
\begin{equation}
  \label{eq:cut.1}
  |y(t)|, |y(b)|, |y(h)| < 2.5,
\end{equation}
reflecting the tracking acceptance and, consequently, the geometric
range in which $b$-tagging is operational. For the processes in
Eqs.~(\ref{eq:top.pair}) and (\ref{eq:top.pair.higgs}), the cuts in
Eq.~(\ref{eq:cut.1}) are the only ones applied. For the process in
Eq.~(\ref{eq:top.pair.nunu}), we apply Eq.~(\ref{eq:cut.1}) together
with an invariant mass cut on the neutrino pair:
\begin{equation}
  \label{eq:cut.2}
  m(\nu,\bar{\nu}) > 150~\mathrm{GeV},
\end{equation}
which ensures that the neutrino pair remains away from the $Z$-boson resonance.

In the case of the processes in Eqs.~(\ref{eq:top.pair.mumu}),
(\ref{eq:sngl.top}), and (\ref{eq:sngl.top.h}), which feature muons in
the final state, we define three distinct rapidity regions:
\begin{equation}
  \label{eq:cut.regions}
  R_{<y_0}:\;|y(\mu)|<y_0,
  \quad \text{with }
 y_0=2.5,\,4.0,\,6.0.
\end{equation}
The phase-space region \RI\ covers the tracking facilities of the
baseline central detector,\footnote{As described in the
  \textsc{Delphes}~\cite{fav14} detector card
  \texttt{Delphes\_card\_MuonColliderDet.tcl}.}  while regions \RII\
and \RIII\ correspond to hypothetical extensions of the muon detector
into the forward and backward hemispheres. We also define the region
\RIc\ for processes where the final-state muon falls outside the
assumed detector acceptance and is therefore unobserved.  For the
process in Eq.~(\ref{eq:top.pair.mumu}) within regions \RI\ and \RII,
we apply the cuts in Eq.~(\ref{eq:cut.1}) alongside:
\begin{equation}
  \label{eq:cut.3}
  \Delta R(\mu^-,\mu^+)>0.4,
  \qquad
  m(\mu^-,\mu^+)>150~\mathrm{GeV},
\end{equation}
which suppresses contributions from the $Z$ resonance. For
Eq.~(\ref{eq:top.pair.mumu}) in region \RIII, we impose
Eqs.~(\ref{eq:cut.1}) and (\ref{eq:cut.3}) in addition to:
\begin{equation}
  \label{eq:cut.4}
  E(\mu) > 500~\mathrm{GeV}.
\end{equation}
Finally, for Eq.~(\ref{eq:top.pair.mumu}) in region \RIc, we apply the
invariant mass cut from Eq.~(\ref{eq:cut.3}), which acts as a missing
mass constraint in this kinematic configuration.  For the single-top
quark processes in Eqs.~(\ref{eq:sngl.top}) and (\ref{eq:sngl.top.h})
within regions \RI, \RII, and \RIII, we implement the cuts from
Eq.~(\ref{eq:cut.1}) along with:
\begin{equation}
  \label{eq:cut.6}
  m(b,\mu,\nu) > 202~\mathrm{GeV}, 
\end{equation}
which forces a potential leptonically decaying antitop line to remain
off-shell, as appropriate for a dedicated single-top production
process rather than top-pair production.

\section{Limits from Top production processes at the MuC}
\label{sec:limits}

To estimate the potential MuC individual limits for each operator
coefficient, we use the same statistical criterion as in
Ref.~\cite{han24}. The signal observability is given by the
statistical significance
\begin{equation}
  \label{eq:stat.signif}
  \begin{gathered}
    \chi = \frac{S}{\sqrt{B}} \; , \\
    S = {{\cal L} \cdot
      (\sigma_\mathrm{SMEFT}-\sigma_\mathrm{SM}) \cdot K}~,
    \quad
    B = {{\cal L} \cdot \sigma_\mathrm{SM} \cdot K}~,
  \end{gathered}
\end{equation}
where $S$ is the number of signal events, ${\cal L}$ the integrated
luminosity, $K$ an efficiency factor that includes branching fractions
and experimental acceptances, and $B$ the number of SM background
events. In the case of a process with $N_t$ top quarks and $N_H$ Higgs
bosons in the final state (where, in this paper, $N_t=1$, 2 and
$N_H=0$, 1), in the hadronic channel we have
\begin{subequations}
  \label{eq:k.factor}
  \begin{gather}
    \label{eq:k.factor.1}
    K={\rm BR}(t\to Wb)^{N_t}{\rm BR}(W\to jj)^{N_t} (\toptag)^{N_t}
    {\rm BR}(h\to b\bar b)^{N_H} (\higgstag)^{N_H},
    \\
    \intertext{where}
    \label{eq:k.factor.2}
    \begin{gathered}
    {\rm BR}(t\to Wb) =0.997, 
    \;\; {\rm BR}(W\to jj)=0.68,
    \;\; \toptag=0.81 \; ,\\
    \;\; {\rm BR}(h\to b\bar b)=0.53,
    \;\; \higgstag=0.75 \; .
    \end{gathered}    
  \end{gather}
\end{subequations}
For instance, in the case of $t\bar{t}$ production, we obtain $K=0.30$.
In (\ref{eq:k.factor.2}), the branching ratios are as given in
\cite{PDG}, and the tagging efficiencies are taken from \cite{CMSDP}
(see also eq.\ (22) in \cite{han24}).  In what follows we will
consider a statistical significance $\chi =2$ in
(\ref{eq:stat.signif}), for $95\%$ CL limits, as in Ref.~\cite{han24}.
Given the luminosities in Table~\ref{tab:lumin} and the SM cross
sections in Table~\ref{tab:incertezas}, we obtain the associated
$1\sigma$ cross-section uncertainties from (\ref{eq:stat.signif}) as
$\varepsilon_\sigma = \Delta \sigma / \sigma_\mathrm{SM} =
(\chi/\sqrt{B})|_{\chi=1} = 1/\sqrt{{\cal L} \cdot \sigma_\mathrm{SM}
  \cdot K}$, for each process and energy.  For clarity, in
Table~\ref{tab:incertezas} we include also the numerical results for
the $2\sigma$ uncertainties.

\begin{table} \centering{}  
\begin{tabular}{|r|| r| r|| r| r|| r| r|| r| r| }
\hline
 \rule{0pt}{11pt}
$\mu^+ \mu^- \rightarrow X$ & \multicolumn{2}{c||}{3 TeV} & \multicolumn{2}{c||}{6 TeV} & \multicolumn{2}{c||}{10 TeV} & \multicolumn{2}{c|}{14 TeV}   \\
\cline{2-9}
 \rule{0pt}{11pt}
 & $\sigma_\mathrm{SM}$& $2\varepsilon_\sigma$& $\sigma_\mathrm{SM}$& $2\varepsilon_\sigma$& $\sigma_\mathrm{SM}$& $2\varepsilon_\sigma$& $\sigma_\mathrm{SM}$& $2\varepsilon_\sigma$\\[2pt]  
\hline
\hline 
 \rule{0pt}{11pt}
$t\bar{t}h$ & 0.40 & 30.5 & 0.13 & 27.1 & 0.053 & 25.2 & 0.030 & 24.0 \\
\hline 
 \rule{0pt}{11pt}
$t\bar{t}$ & 19.33 & 2.8 & 4.84 & 2.8 & 1.75 & 2.8 & 0.89 & 2.8 \\
\hline 
 \rule{0pt}{11pt}
$t\bar{b}\mu^-\bar{\nu}_\mu \ [R_{<2.5}]$ & 2.46 & 5.7 & 1.58 & 3.6 & 1.01 & 2.7 & 0.71 & 2.3 \\
 \rule{0pt}{11pt}
$t\bar{b}\mu^-\bar{\nu}_\mu \ [R_{<4.0}]$ & 5.28 & 3.9 & 5.46 & 1.9 & 4.82 & 1.2 & 4.14 & 0.9 \\
 \rule{0pt}{11pt}
$t\bar{b}\mu^-\bar{\nu}_\mu \ [R_{<6.0}]$ & 6.75 & 3.5 & 9.32 & 1.5 & 10.38 & 0.8 & 10.68 & 0.6 \\
\hline 
 \rule{0pt}{11pt}
$t\bar{t}\mu^+\mu^- \ [R_{<2.5}]$ & 0.093 & 39.9 & 0.043 & 29.3 & 0.021 & 25.1 & 0.013 & 22.7 \\
 \rule{0pt}{11pt}
$t\bar{t}\mu^+\mu^- \ [R_{<4.0}]$ & 0.90 & 12.8 & 0.67 & 7.4 & 0.37 & 6.0 & 0.22 & 5.5 \\
 \rule{0pt}{11pt}
$t\bar{t}\mu^+\mu^- \ [R_{<6.0}]$ & 2.05 & 8.5 & 3.00 & 3.5 & 3.00 & 2.1 & 2.77 & 1.6 \\
 \rule{0pt}{11pt}
$t\bar{t}\mu^+\mu^- \ [R_{>2.5}]$ & 7.03 & 4.6 & 13.80 & 1.6 & 18.65 & 0.8 & 21.52 & 0.6 \\
\hline 
 \rule{0pt}{11pt}
$t\bar{t}\nu\bar{\nu}$ & 3.86 & 6.2 & 7.29 & 2.2 & 9.68 & 1.2 & 11.02 & 0.8 \\
\hline 
 \rule{0pt}{11pt}
$t\bar{t}(\mu^+\mu^- + \nu\bar{\nu})$ & 10.89 & 3.7 & 21.09 & 1.3 & 28.33 & 0.7 & 32.54 & 0.5 \\
\hline 
 \rule{0pt}{11pt}
$t\bar{b}\mu^-\bar{\nu}_\mu h \ [R_{<2.5}]$ & 0.012 & 129.8 & 0.0086 & 77.1 & 0.0062 & 54.5 & 0.0047 & 44.4 \\
 \rule{0pt}{11pt}
$t\bar{b}\mu^-\bar{\nu}_\mu h \ [R_{<4.0}]$ & 0.019 & 103.8 & 0.021 & 48.7 & 0.021 & 29.8 & 0.019 & 22.3 \\
 \rule{0pt}{11pt}
$t\bar{b}\mu^-\bar{\nu}_\mu h \ [R_{<6.0}]$ & 0.014 & 119.4 & 0.031 & 40.3 & 0.043 & 20.7 & 0.048 & 13.9 \\
\hline
\end{tabular}

\caption{SM cross sections $\sigma_\mathrm{SM}$ in fb, computed with
  MG5, and their $2\sigma$ relative uncertainties
  $2\varepsilon_\sigma$ in \%.} 
  \label{tab:incertezas}
  \end{table}

In the following subsections we address each one of the top
production processes and the individual limits obtained for each WC.
Given a certain operator, there are usually one or two processes
that yield the most sensitivity.  On the other hand, given a certain
process, there are some operators that are particularly sensitive to
it and others that are not.  To simplify the presentation, not all
the WC bounds coming from a certain process are shown, but only the
ones for which the best limits are obtained.
The limits from the five most sensitive processes for each
coefficient and energy are collected in Appendix~\ref{app:limits},
except for $C_{bW}$ (see Sect.~\ref{sec:limsinglet}).  The Appendix
also shows the effect of an assumed $2\%$ systematic uncertainty,
which can be significant for processes with large cross sections, and
hence small statistical uncertainties.


\subsection{Limits from $s$-channel $t\bar t$ production}
\label{sec:limtt}

We first consider $t\bar t$ production. As we go from $\sqs = 3$~TeV
to $14$~TeV, the SM cross section is reduced by an order of magnitude,
but the luminosity also grows by an order of magnitude, so that the
$2\sigma$ uncertainty remains constant at $2.8\%$, see
Table~\ref{tab:incertezas}. Only two of the four coefficients that
can be constrained from this process, $C^{(-)}_{\vphi Q}$,
$C_{\vphi t}$, $C_{tB}$, and $C_{tW}$, get significant limits: the
dipole coefficients $C_{tB}$ and $C_{tW}$.  The $ttZ$ and $tt\gamma$
couplings from the dipole operators $Q_{uB}$ and $Q_{uW}$ are
proportional to the $Z$-boson and photon momenta; thus, not
surprisingly, some of the best limits for $C_{tB}$ and $C_{tW}$ at
high energy come from this process. The $ttZ$ couplings from
$Q^{(-)}_{\vphi Q}$ and $Q_{\vphi t}$ are not momentum-dependent, and
the competitive limits do not come from this process.  For
$C_{\vphi t}$, the limits are greater than the current ones shown in
Table~\ref{tab:cmsatlaslim}.  For $C^{(-)}_{\vphi Q}$, the limits are
of order $1$ (a third of the current bounds), yet they are about an
order of magnitude greater than what we obtain from $t\bar t$
production with neutrinos.  That is the subject of section
\ref{sec:limttnn}.

Regarding previous reports on limits on the dipole coefficients from
$s$-channel $t\bar t$ production, other authors have also found that
this process yields competitive bounds.
In Table~\ref{tab:ctwctbfromtt} we show our limits as well as the
limits presented in Refs.~\cite{miralles2503,han24}.
Ref.~\cite{han24} provides one lower limit on $\Lambda/\sqrt{C_{tB}}$
and on $\Lambda/\sqrt{C_{tW}}$, and we show the corresponding
upper limit on $C/\Lambda^2$ with $\Lambda = 1$~TeV.
The authors in \cite{miralles2503} also provide only one upper bound.
Since we are using the same criterion for the cross-section
uncertainties as Ref.~\cite{han24}, it is expected that our
limits should coincide with theirs.  However, in \cite{han24} the
authors include a correction from initial-state radiation (ISR),
which tends to increase the $t\bar t$ cross section by a small amount.
This is why their limits are also slightly smaller.
The coefficients $C_{tB}$ and $C_{tW}$ are, in general, complex.
We provide the limits on the imaginary parts of all the WC in the
Appendix.  In \cite{han24}, there are limits for $C_{tBi}$ given at
$\sqs = 3, 10$~TeV, and the bounds we obtain are again consistent with
theirs.
For the real parts at $3$~TeV, there is a second allowed interval
along each axis, for both $C_{tWr}$ and $C_{tBr}$ (see
Fig.~\ref{fig:contours_r}); the tabulated intervals are those
containing the SM point.

\begin{table} \centering{}
  \begin{tabular}{|c||c|c|c|c|c|}
    \cline{1-6} $ $ & $C^{}_{tWr}$ & $C^{}_{tBr}$ &
    $C^{}_{tWr}$ \cite{han24} & $C^{}_{tBr}$ \cite{han24} &
    $C^{}_{tWr}$ \cite{miralles2503}\\
    \hline
    $t\bar t-3$ TeV & 
    $[-0.029 ,+0.053]$ & $[-0.023 ,+0.038]$ &
$0.026$ & $0.022$ & $0.028$ \\
    \hline
    $t\bar t-6$ TeV &
    $[-0.022 ,+0.062]$ & $[-0.018 ,+0.051]$ &
$--$ & $--$ & $--$ \\ 
    \hline
    $t\bar t-10$ TeV &
    $[-0.016 ,+0.031]$ & $[-0.013 ,+0.025]$ &
    $0.016$ & $0.013$ & $0.015$ \\
    \hline
    $t\bar t-14$ TeV &
    $[-0.013 ,+0.020]$ & $[-0.010 ,+0.016]$ &
$--$ & $--$ & $--$ \\ 
    \hline 
  \end{tabular}
  \caption{Individual limits on the dipole coefficients
    $C_{tWr}$ and $C_{tBr}$ from $\mu^+ \mu^- \to t\bar t$.
    The limits correspond to
    $2\varepsilon_\sigma = 2.8\%$ for all values of $\sqrt{s}$.}
  \label{tab:ctwctbfromtt}
  \end{table}

\subsection{Top pair plus Higgs production $\mu^+ \mu^- \to t{\bar t} h$}
\label{sec:limtth}

In Fig.~\ref{fig:dgrms.1}(b) and (c), we show two sample diagrams of
the $s$-channel $t{\bar t} h$ production mode.
In addition to the potential contributions from the same four
couplings as in the previous process, the Yukawa coupling also
contributes, as suggested by diagram (b):
\begin{equation}                                                            
y_t = 1 + \delta_{yt} =                                                     
1-\frac{v}{\sqrt{2} m_t} \left(\frac{v}{\Lambda}\right)^2                   
C_{t\vphi} = 1-\frac{C_{t\vphi}}{16.4}\; .                                  
\end{equation}
It is natural to expect that this production mode is the most
sensitive to the top-Higgs Yukawa coupling.
In 2022, the authors of Ref.~\cite{meade22} obtained bounds on
$\delta_{yt}$ based on the $s$-channel $t\bar t h$ production at
$3$~TeV and also based on VBF $t\bar t h$ production at $10$~TeV.
Their study included a detailed analysis of signal versus
background for both cases, and they found that the precision for
the Yukawa measurement would be at best about $\sim 50\%$.  This is
equivalent to an upper bound on $C_{t\vphi}$ of about $8.2$.
At $\sqs = 10$~TeV, we obtain bounds in the interval
$[-2.5, +2.9]$, significantly stronger than the $8.2$ bound above,
but still somewhat weaker than the current limits from ATLAS
listed in Table~\ref{tab:cmsatlaslim}.
In subsequent articles~\cite{miranda22,wangyukawa}, it was reported
that the process most sensitive to $\delta_{yt}$ is
$t{\bar t} \nu_\mu \bar \nu_\mu$ production.  In this case the
attainable precision improves by about an order of magnitude.
We will refer to this process in the next section.

Surprisingly, at high energies the $s$-channel $t\bar t h$ process
provides the most stringent bounds on the dipole operators $Q_{tB}$
and $Q_{tW}$. This occurs despite the lower SM cross section and
the resulting large cross-section uncertainty
(see Table~\ref{tab:incertezas}) compared to $t\bar t$ production
\cite{han24}.
We note that Ref.~\cite{mele26} also considers this mode (along
with $t\bar t$) to obtain limits on four-fermion couplings.
In Table~\ref{tab:ctwctbfromtth}, we show the limits on $C_{tB}$
and $C_{tW}$ obtained from this process.
The limits given in Refs.~\cite{han24,miralles2503} are also
shown.  As in the case of $t\bar t$, our limits agree with those
of Ref.~\cite{han24}.  This is also true for the imaginary part of
$C_{tB}$. In fact, at all energies the best limits on the
imaginary parts $C_{tBi}$ and $C_{tWi}$ come from $s$-channel
$t\bar t h$ production.
Nevertheless, as we will see below, there are other processes,
such as single-top and single-top-plus-Higgs production, that
yield less stringent but still competitive bounds on the dipole
coefficients.

\begin{table} \centering{}
  \begin{tabular}{|c|c|c|c|c|c|}
    \cline{1-6} $ $ & $C^{}_{tWr}$ & $C^{}_{tBr}$ &
    $C^{}_{tWr}$ \cite{han24} & $C^{}_{tBr}$ \cite{han24} &
 $C^{}_{tWr}$ \cite{miralles2503}\\
    \hline
    $t\bar th-3$ TeV & 
    $[-0.028 ,+0.134]$ & $[-0.022 ,+0.112]$ &
    $0.029$ & $0.023$ & $0.03$ \\
    \hline
    $t\bar th-6$ TeV & 
    $[-0.009 ,+0.036]$ & $[-0.007 ,+0.030]$ &
$--$ & $--$ & $--$ \\    
    \hline
    $t\bar th-10$ TeV & 
    $[-0.004 ,+0.013]$ & $[-0.003 ,+0.011]$ &
$0.004$ & $0.003$ & $0.007$ \\
    \hline
    $t\bar th-14$ TeV & 
    $[-0.002 ,+0.007]$ & $[-0.002 ,+0.006]$ &
$--$ & $--$ & $--$ \\    
    \hline 
  \end{tabular}
  \caption{Individual limits on the dipole coefficients
    $C_{tWr}$ and $C_{tBr}$ from $\mu^+ \mu^- \to t\bar t h$.
The limits correspond to the cross-section uncertainties
    given in Table~\ref{tab:incertezas}.}
  \label{tab:ctwctbfromtth}
  \end{table}

  
\subsection{The process
$\mu^+ \mu^- \to t{\bar t} \nu \bar \nu$}
\label{sec:limttnn}

In the effective $W$ approximation (EWA), this process originates from
the VBF subprocess $W^+W^- \to t\bar t$.  It is well known that, in
addition, the $ZZ(\gamma) \to t\bar t$ contribution must be included
for events in which the final-state muons are not detected.  However,
before addressing the additional contribution, it is instructive to
see what sensitivity we obtain for the $W^+W^-$ mode alone.  As
mentioned in the previous section, this process is more sensitive to
the Yukawa coupling than $t\bar t h$ production at $\sqs \geq 6$~TeV,
and for the imaginary part at all energies.  The limit
$|C_{t\vphi}| \lesssim 1$ in Table~\ref{tab:limttnunu} at $10$~TeV is
equivalent to a variation of about $6.5\%$ for the Yukawa
coefficient. This is consistent with the limit of $5.6\%$ for
$\delta_{yt}$ given in Ref.~\cite{miranda22}, where the enhanced
sensitivity comes from analyzing the distribution of the scattering
angle between the top and the $W^+$ boson.

In a subsequent study, Ref.~\cite{wangyukawa} reports an even
smaller $95\%$ CL limit of $3\%$ based on two kinematic
distributions: the $t\bar t$ invariant mass and the polar angle of
$t$.  Furthermore, Ref.~\cite{wangyukawa} considers both the
semileptonic and the hadronic decay modes of the top-quark pair,
whereas Ref.~\cite{miranda22} uses only the semileptonic mode.
In the approach of Ref.~\cite{wangyukawa}, the partonic cross
sections for both processes $W^+W^- \to t\bar t$ and
$ZZ(\gamma) \to t\bar t$ are calculated and then convolved with the
muon PDF \cite{muonsmasher}.
The limit from Ref.~\cite{wangyukawa} translates to
$|C_{t\vphi}| \lesssim 0.5$, which is about half of what we obtain.
The authors of \cite{wangyukawa} point out that the sensitivity is
dominated by the $W^+W^-$ mode, a result that is also consistent
with the weaker limits we obtain from $t\bar t \mu^+\mu^-$
production, as we shall see below.

More recently, Ref.~\cite{chen25} studied the Yukawa coupling in the
Higgs-associated production process
$\mu^+\mu^- \to t\bar t \nu_\mu \bar\nu_\mu h$, obtaining a $2\sigma$
limit of $4.9\%$ on $\delta_{yt}$ at $\sqs = 10$~TeV.  Their analysis
is based on the full matrix element but without including top decay
(as we do here).  However, they also re-evaluate this limit including
top decay in the leptonic and semileptonic channels (at the partonic
level) and obtain a weaker bound that is twice as large.

As for the imaginary part $C_{t\vphi i}$ of the Yukawa coupling, the
process $t{\bar t} \nu \bar \nu$ provides a bound of about
$1.2$ at $\sqs = 10$~TeV (see Fig.~\ref{fig:limits_2}), which increases
to $1.5$ when the untagged-muon contribution is added.  In
Ref.~\cite{cpyukawa23}, a study based on the
$t{\bar t} \nu_\mu \bar \nu_\mu h$ and single-top-plus-Higgs final
states uses a CP-violation-sensitive angular variable that directly
tests the imaginary coefficient.  The analysis includes event
reconstruction and signal-versus-background separation, and the
authors obtain a $2\sigma$ limit of $|C_{t\vphi i}| \leq 4$, weaker
than the one we obtain.

Besides the Yukawa coefficient, we have found that the vector left-
and right-handed $ttZ$ couplings $C^{(-)}_{\vphi Q}$ and $C_{\vphi t}$
also get their best bounds from the $t{\bar t} \nu \bar \nu$
process. Notice that for $C_{\vphi t}$ at $10$~TeV, the bound is more
than an order of magnitude (about a factor of 40) smaller than the
current LHC limits. As we will see in the next section, when the
contribution from $t\bar t\mu^+\mu^-$ production with untagged forward
muons is added, the sensitivity to $C^{(-)}_{\vphi Q}$ barely changes,
but the limits on $C_{\vphi t}$ become weaker by factors of about $3$
to $7$, and those on $C_{t\varphi}$ by about $30\%$.

\begin{table} \centering{}
  \begin{tabular}{|c||c|c|c|}
    \cline{1-4} $ $ & 
 $C^{(-)}_{\vphi Q}$ & $C^{}_{\vphi t}$ & $C^{}_{t \vphi r}$ \\
    \hline
$3$ TeV & $[-0.52 ,+0.55]$ & $[-0.38 ,+0.48]$ & $[-3.47 ,+4.49]$ \\
    \hline
$6$ TeV & $[-0.18 ,+0.19]$ & $[-0.11 ,+0.12]$ & $[-1.67 ,+2.41]$ \\
    \hline
$10$ TeV & $[-0.10 ,+0.10]$ & $[-0.050 ,+0.056]$ & $[-1.07 ,+1.68]$ \\
    \hline
$14$ TeV & $[-0.063 ,+0.065]$ & $[-0.030 ,+0.034]$ & $[-0.79 ,+1.32]$ \\
    \hline
  \end{tabular}
  \caption{Individual limits on coefficients $C^{(-)}_{\vphi Q}$,
    $C_{\vphi t}$ and $C_{t \vphi r}$ from the production mode
    $t\bar t \nu \bar \nu$.  The limits correspond to the
    cross-section uncertainties given in Table~\ref{tab:incertezas}.}
  \label{tab:limttnunu}
  \end{table}

\subsection{The process
$\mu^+ \mu^- \to t{\bar t} \mu^+ \mu^-$}
\label{sec:limttmm}

We have found that this process, which involves tagging the
final-state muons, is not very sensitive to any of the couplings
considered here.  At most, we obtain second-best limits for
$C_{\vphi t}$, and among the best five for $C^{(-)}_{\vphi Q}$. Even
then, those bounds are significantly weaker than the ones from the
previous case $\mu^+\mu^- \to t{\bar t} \nu \bar \nu$.  Moreover, the
requirement that the outgoing muons appear in the central region
$|y| < 2.5$ greatly reduces the sensitivity.  Only if we allow for
detection at high rapidities do we obtain bounds for
$C^{(-)}_{\vphi Q}$ and $C_{\vphi t}$ stronger than the current LHC
limits.  The limits for $|y| < 2.5$ are either at least three times
larger than the current LHC bounds or about an order of magnitude
larger than the limits for $|y| < 4$ or $|y| < 6$. Since they are not
competitive, they are omitted from Table~\ref{tab:limttmumu}.

Although this process is less sensitive, it must be added to the
$t{\bar t} \nu \bar \nu$ signal when the outgoing muons escape
detection.  Assuming a muon detector that covers the region
$|y(\mu^\pm)| < 2.5$, we have re-evaluated the limits from
$t{\bar t} \nu \bar \nu$ by including $t{\bar t} \mu^+ \mu^-$ with
both muons satisfying $|y(\mu^\pm)| > 2.5$.  We show the new limits
obtained from this process, which is dominated by VBF, in
Table~\ref{tab:limttnunumumu}. Comparing with the $10$~TeV results in
Table~\ref{tab:limttnunu}, we observe that the sensitivity to
$C^{(-)}_{\vphi Q}$ is essentially unchanged, while the limits on
$C_{\vphi t}$ and $C_{t\varphi}$ become weaker by factors of about $3$
and $1.3$, respectively.

\begin{table} \centering{}
  \begin{tabular}{|c||c|c|c|}
    \cline{1-4} $ $ & 
 $C^{(-)}_{\vphi Q}$ & $C^{}_{\vphi t}$ & $C^{}_{t \vphi r}$ \\
    \hline
 $3$ TeV $ [R_{<4.0}]$ &
    $[-0.73 ,+1.01]$ & $[-3.30 ,+1.91]$ & $[-7.34 ,+15.1]$ \\
$3$ TeV $ [R_{<6.0}]$ &
    $[-1.16 ,+1.36]$ & $[-1.74 ,+1.40]$ & $[-8.01 ,+16.1]$ \\
    \hline
    $6$ TeV $ [R_{<4.0}]$ &
    $[-0.41 ,+0.47]$ & $[-1.81 ,+1.66]$ & $[-4.51 ,+11.4]$ \\
    $6$ TeV $ [R_{<6.0}]$ &
    $[-0.46 ,+0.50]$ & $[-0.70 ,+0.59]$ & $[-3.02 ,+8.78]$ \\
    \hline
    $10$ TeV $ [R_{<4.0}]$ &
    $[-0.35 ,+0.41]$ & $[-2.16 ,+1.31]$ & $[-4.59 ,+11.80]$ \\
   $10$ TeV $ [R_{<6.0}]$ &
    $[-0.26 ,+0.27]$ & $[-0.41 ,+0.35]$ & $[-1.62 ,+6.34]$ \\
    \hline
    $14$ TeV $ [R_{<4.0}]$ &
    $[-0.31 ,+0.36]$ & $[-2.26 ,+1.40]$ & $[-5.35 ,+13.2]$ \\
    $14$ TeV $ [R_{<6.0}]$ &
    $[-0.19 ,+0.20]$ & $[-0.30 ,+0.27]$ & $[-1.16 ,+5.38]$ \\   
    \hline
  \end{tabular}
  \caption{Individual limits on coefficients
$C^{(-)}_{\vphi Q}$, $C_{\vphi t}$ and $C_{t \vphi r}$ from the
    production mode $t\bar t \mu^+ \mu^-$.
The limits correspond to the cross-section uncertainties
    given in Table~\ref{tab:incertezas}.}
  \label{tab:limttmumu}
  \end{table}

\begin{table} \centering{}
  \begin{tabular}{|r||c|c|c|}
    \cline{1-4} $ $ & 
 $C^{(-)}_{\vphi Q}$ & $C^{}_{\vphi t}$ & $C^{}_{t \vphi r}$ \\
    \hline
$3$ TeV & $[-0.54 ,+0.58]$ & $[-0.86 ,+2.71]$ & $[-4.30 ,+5.69]$ \\
    \hline
$6$ TeV & $[-0.19 ,+0.19]$ & $[-0.25 ,+1.29]$ & $[-2.01 ,+3.14]$ \\
    \hline
$10$ TeV& $[-0.10 ,+0.10]$ & $[-0.11 ,+0.20]$ & $[-1.26 ,+2.23]$ \\
    \hline
$14$ TeV& $[-0.070 ,+0.070]$ & $[-0.070 ,+0.12]$ & $[-0.97 ,+1.81]$ \\
    \hline
  \end{tabular}
  \caption{Individual limits on coefficients
$C^{(-)}_{\vphi Q}$, $C_{\vphi t}$ and $C_{t \vphi r}$ from the
process $t\bar t \nu \bar \nu + t\bar t \mu^+ \mu^-$ with $|y (\mu^\pm)|> 2.5$. 
The limits correspond to the cross-section uncertainties given in
Table~\ref{tab:incertezas}.} 
  \label{tab:limttnunumumu}
  \end{table}

\subsection{Single-top production
  $\mu^+ \mu^- \to t{\bar b} \mu^- \bar \nu_\mu +
 {\bar t}b \mu^+ \nu_\mu$}
\label{sec:limsinglet}

Single-top production has been considered for an $e^+ e^-$ collider at
energies of $0.5$, $1$, and $3$~TeV \cite{escamilla}, but it has not
been addressed in the literature for the higher energies reached by
the MuC.  In our analysis, we observe that single-top production
provides the strongest limits on the right-handed $tbW$ coupling
$C_{\vphi tb}$.  The limits on the dipole couplings $C_{tB}$, $C_{tW}$
and $C_{bW}$ are not the strongest, but they improve on the current
LHC limits in Table~\ref{tab:cmsatlaslim}: for $C_{bW}$ at all
energies, and for $C_{tB}$ and $C_{tW}$ from $6$~TeV on, with
intervals narrower by factors of about $2$--$3$ at $6$~TeV and
$5$--$13$ at $10$ and $14$~TeV.  Our results are shown in
Table~\ref{tab:limsinglet}.  In Ref.~\cite{escamilla}, at $3$~TeV, the
limits on $C_{\vphi tb}$ were established as $\pm 3.1$ for a
cross-section uncertainty of $5\%$.  This limit is similar to the one
we obtain at $3$~TeV with $|y| < 2.5$ at the MuC.

The limits presented in Table~\ref{tab:limsinglet} depend on the muon
rapidity range. We show the limits for the range that gives the best
result (for $C_{\vphi tb}$ at $3$~TeV, the range $|y| < 4$ is
marginally better; see Fig.~\ref{fig:limits_2}); those for the other
ranges are larger, in some cases only slightly.
Notice that the limits on $C_{tW}$ and $C_{bW}$ are
similar, both below $10^{-1}$.  Both operators contribute to
single-top production via the $tbW$ coupling, with $Q_{tW}$ involving
a left-handed $b$ quark and $Q_{bW}$ a right-handed one.  The
interference with the SM amplitude for $Q_{bW}$ is suppressed by $m_b$
because of the chirality flip. However, the contribution to the
amplitude squared is dominated by the term proportional to $C^2_{bW}$,
which we find to be similar in size to the term proportional to
$C^2_{tW}$.  The right-handed charged-current operator $Q_{\vphi tb}$
lacks the dependence on the $W$-boson momentum that the dipole
operators generate in the $tbW$ vertex, and its contribution to the
amplitude is significantly weaker than that of the dipole
operators. The resulting limits are, as expected, much larger.

A very recent study of the potential sensitivity of the HL-LHC to this
coupling, using $bW$ spin correlations \cite{wang26}, projects a bound
on the right-handed $tbW$ form factor of $|V_{R}| < 0.037$, which
translates to $|C_{\vphi tb}|=2 \Lambda^2/v^2 |V_R| < 1.22$.  If muon
tagging extends to $|y| < 6$, the MuC at $10$~TeV could set a bound
about a third of that value, as seen in Table~\ref{tab:limsinglet}.
For $|y| < 2.5$, the bound at that energy would be about $1.1$, which
is similar to their prediction.

For the bottom-quark dipole operator $Q_{bW}$, the strongest
limits on $C_{bW}$ come from $b\bar b$ and $b\bar b h$ production.
A recent study, Ref.~\cite{barducci}, provides $95\%$ CL limits on
$C_{bW}$ and $C_{bB}$ of order $10^{-2}$ or smaller, as is the case
for the dipole coefficients $C_{tW}$ and $C_{tB}$ with
$t\bar t h$ production.  For $C_{bW}$, they read:
\begin{alignat*}{2}
  -0.030 &\leq C_{bWr} \leq +0.028 &\qquad& (\sqs = 3~{\rm TeV}) \; , \nonumber \\
  -0.0024 &\leq C_{bWr} \leq +0.0020 && (\sqs = 10~{\rm TeV}) \; .
\end{alignat*}

\begin{table} \centering{}
  \begin{tabular}{|c||c|c|c|c|}
    \cline{1-5} $ $ & 
    $C_{\vphi tbr}$,$|y|<6.0$ & $C^{}_{tBr}$,$|y|<2.5$
    & $C^{}_{tWr}$,$|y|<2.5$  & $C^{}_{bWr}$,$|y|<2.5$ \\
    \hline
    $3$ TeV & $[-2.14 ,+2.11]$ & $[-0.097 ,+0.15]$
    & $[-0.080 ,+0.21]$ & $[-0.13 ,+0.14]$ \\
    \hline
    $6$ TeV & $[-0.76 ,+0.69]$ & $[-0.038 ,+0.041]$ &
    $[-0.034 ,+0.059]$ & $[-0.046 ,+0.047]$ \\
    \hline
    $10$ TeV & $[-0.35 ,+0.31]$ & $[-0.017 ,+0.016]$
    & $[-0.016 ,+0.023]$ & $[-0.020 ,+0.021]$ \\
    \hline
    $14$ TeV & $[-0.22 ,+0.20]$ & $[-0.010 ,+0.009]$
    & $[-0.010 ,+0.012]$ & $[-0.011 ,+0.011]$ \\
    \hline
  \end{tabular}
  \caption{Individual limits on the coefficients $C_{\vphi tbr}$,
    $C_{tBr}$, $C_{tWr}$, and $C_{bWr}$ from the single-top production
    modes $t\bar b \mu^- \bar \nu_\mu + \bar t b \mu^+ \nu_\mu $.  The
    range of $|y(\mu^\pm)|$ shown is the one that gives the best
    limits (some of the other rapidity ranges are shown in
    Figs.~\ref{fig:limits_1} and~\ref{fig:limits_2}).  The limits
    correspond to the cross-section uncertainties given in
    Table~\ref{tab:incertezas}.}
  \label{tab:limsinglet}
  \end{table}


\subsection{Single-top production plus Higgs}
\label{sec:limsingleth}

The bounds on WCs obtained from single-top-plus-Higgs production are
shown in Table~\ref{tab:limsingleth}.  They show the same relative
hierarchy of bounds as in single-top production (see
Table~\ref{tab:limsinglet}), and the bounds themselves are 
similar for both processes.  However, the nature of those bounds is
different: the single-top process benefits from relatively large
cross sections and low backgrounds, resulting in small
cross-section uncertainties, as shown in Table~\ref{tab:incertezas},
whereas the single-top-plus-Higgs cross section depends on the WC
much more strongly than that of any other process except
top-pair-plus-Higgs production, but, being small, leads to large
statistical uncertainties.  These two effects partially cancel, so
that single-top-plus-Higgs production ends up setting limits similar
to those from single-top production, as shown in
Tables~\ref{tab:limsinglet} and~\ref{tab:limsingleth}.
Consequently, if a systematic uncertainty were added in quadrature to
the statistical ones in Table~\ref{tab:incertezas}, the total
uncertainty for single-top production would increase, weakening its
limits on the WCs, whereas the total uncertainty for
single-top-plus-Higgs, being statistically dominated, would remain
essentially unchanged, and so would the limits in
Table~\ref{tab:limsingleth}.
For instance, for single-top production at $14$~TeV with
$|y| < 6$, a $2\%$ systematic uncertainty would increase the total
$2\sigma$ uncertainty from $0.6\%$ to $4.0\%$, whereas for
single-top-plus-Higgs production it would merely change from $13.9\%$ to
$14.5\%$.  The limits for an assumed systematic uncertainty of $2\%$
are shown in Figs.~\ref{fig:limits_1}--\ref{fig:limits_3} of
Appendix~\ref{app:limits}.

\begin{table} \centering{}
  \begin{tabular}{|c||c|c|c|c|}
    \cline{1-5} $ $ & 
    $C_{\vphi tbr}$,$|y|<6.0$ & $C^{}_{tBr}$,$|y|<2.5$
    & $C^{}_{tWr}$,$|y|<2.5$  & $C^{}_{bWr}$,$|y|<2.5$ \\
    \hline
    $3$ TeV & $[-4.29 ,+4.20]$ & $[-0.14 ,+0.22]$
    & $[-0.13 ,+0.24]$ & $[-0.20 ,+0.20]$ \\
    \hline
    $6$ TeV & $[-1.21 ,+1.18]$ & $[-0.043 ,+0.061]$ &
    $[-0.039 ,+0.063]$ & $[-0.052 ,+0.053]$ \\
    \hline
    $10$ TeV & $[-0.51 ,+0.50]$ & $[-0.017 ,+0.023]$
    & $[-0.015 ,+0.023]$ & $[-0.020 ,+0.020]$ \\
    \hline
    $14$ TeV & $[-0.29 ,+0.28]$ & $[-0.009 ,+0.012]$
    & $[-0.008 ,+0.012]$ & $[-0.010 ,+0.010]$ \\
    \hline
  \end{tabular}
  \caption{Individual limits on the coefficients $C_{\vphi tbr}$,
    $C_{tBr}$, $C_{tWr}$, and $C_{bWr}$ from the single-top production
    plus Higgs modes
    $t\bar b \mu^- \bar \nu_\mu h + \bar t b \mu^+ \nu_\mu h$.  The
    range of $|y(\mu^\pm)|$ shown is the one that gives the best
    limits (some of the other rapidity ranges are shown in
    Figs.~\ref{fig:limits_1} and~\ref{fig:limits_2}).  The limits
    correspond to the cross-section uncertainties given in
    Table~\ref{tab:incertezas}.}
  
  \label{tab:limsingleth}
  \end{table}


\subsection{Allowed regions and marginalized limits for
  $\boldsymbol{C_{tB}}$ and $\boldsymbol{C_{tW}}$} 
\label{sec:marginalized}

\begin{figure}[t]
  \centering
  \includegraphics[scale=0.8]{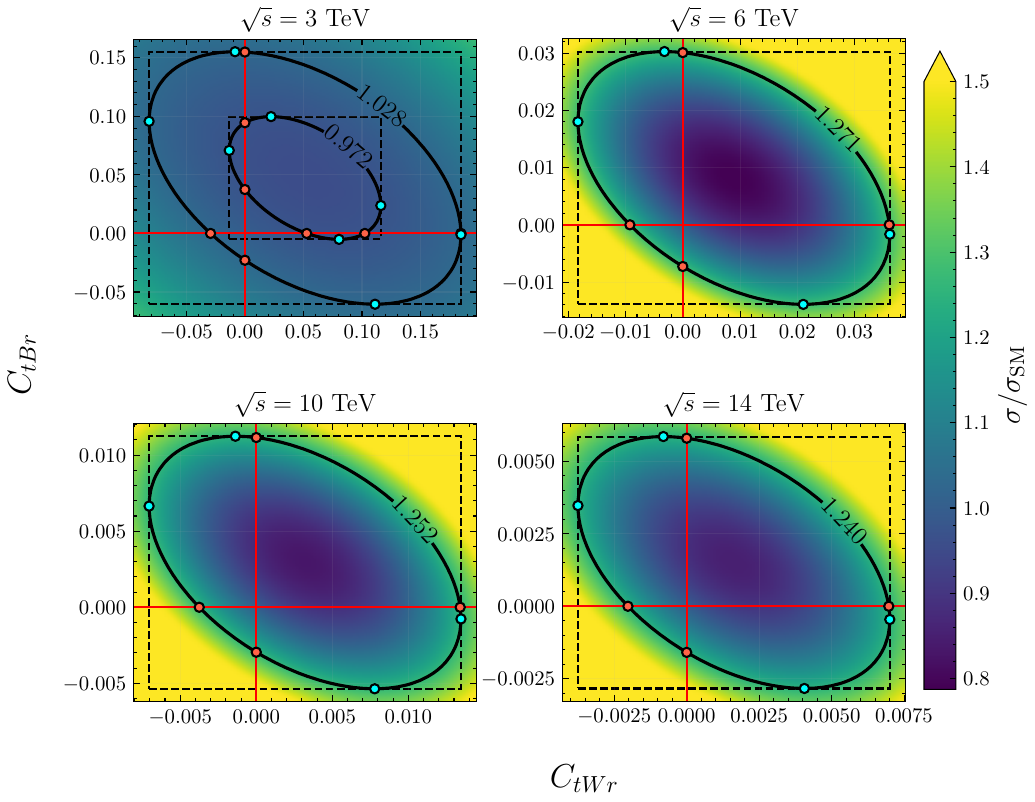}
\caption{Allowed regions at 95\% CL for the real parts of the Wilson
  coefficients $C_{tW}$ and $C_{tB}$. Individual limits are indicated
  by the light-red dots, and marginalized ones by the dashed black lines.}
  \label{fig:contours_r}
\end{figure}

\begin{figure}[t]
  \centering
  \includegraphics[scale=0.8]{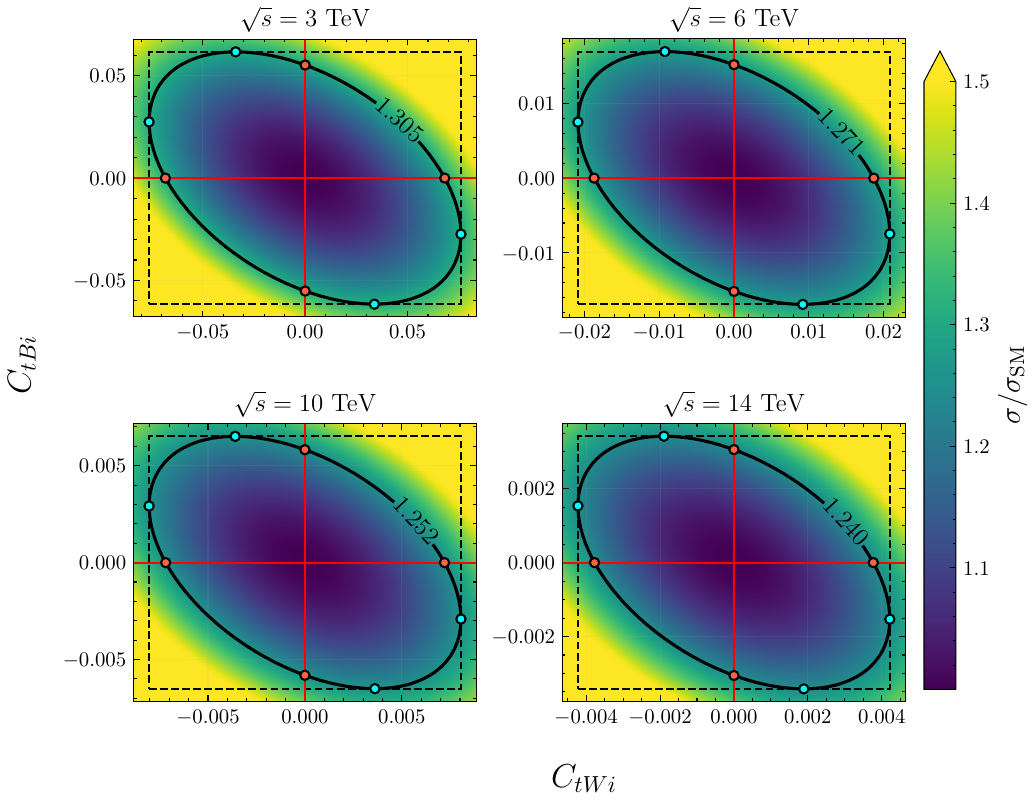}
\caption{Allowed regions at 95\% CL for the imaginary parts of the Wilson
  coefficients $C_{tW}$ and $C_{tB}$. Individual limits are indicated
  by the light-red dots, and marginalized ones by the dashed black lines.}
  \label{fig:contours_i}
\end{figure}

In this section we report two-dimensional allowed regions for the
electroweak dipole Wilson coefficients $C_{tB}$ and $C_{tW}$, the
operators to which we find the strongest sensitivity.  The allowed
regions for the real and imaginary parts are reported separately:
Fig.~\ref{fig:contours_r} shows the allowed regions in the real-parts
plane $C_{tWr}$--$C_{tBr}$, and Fig.~\ref{fig:contours_i} those in the
imaginary-parts plane $C_{tWi}$--$C_{tBi}$, in both cases for the four
energies $\sqs = 3$, $6$, $10$, and $14$~TeV.  The allowed regions at
$2\sigma$ confidence level are bounded by ellipses defined by
$\sigma(C_{tW},C_{tB})/\sigma_\mathrm{SM} = 1 \pm 2
\varepsilon_\sigma$, where $\varepsilon_\sigma$ is the cross-section
uncertainty from Table~\ref{tab:incertezas}, as explicitly indicated
on the ellipses in the figures.  For the real parts at $\sqs = 3$~TeV,
the allowed region is an elliptical annulus, whereas for the remaining
energies, and for the imaginary parts at all energies, it is a solid
ellipse.

The allowed regions are computed for the process that yields the
best individual limits at each energy.  For the real parts
(Fig.~\ref{fig:contours_r}), these are the limits from $t\bar t$
production at $\sqs = 3$~TeV, see Table~\ref{tab:ctwctbfromtt} and
Fig.~\ref{fig:limits_1}, and from $t\bar t h$ production at the
three higher energies, see Table~\ref{tab:ctwctbfromtth} and
Fig.~\ref{fig:limits_1}.  For the imaginary parts
(Fig.~\ref{fig:contours_i}), the best limits are set by
$t\bar t h$ production at all energies, see Fig.~\ref{fig:limits_1}.
In the figures, the extremal points of the ellipses, indicated by
light-blue dots, determine the marginalized limits on the couplings,
as shown by the dashed lines.  The intersections of the ellipses with
the coordinate axes, shown by light-red dots, correspond to the
individual limits given in Tables~\ref{tab:ctwctbfromtt}
and~\ref{tab:ctwctbfromtth} and in Fig.~\ref{fig:limits_1}.

The different shapes of the allowed regions can be understood as
follows.  The amplitudes corresponding to the real parts $C_{tWr}$
and $C_{tBr}$ interfere significantly with the SM amplitude, which
produces a noticeable asymmetry between the lower and upper limits in
Tables~\ref{tab:ctwctbfromtt} and~\ref{tab:ctwctbfromtth}.  If the
interference were negligible or zero, as is the case for $C_{tWi}$ and
$C_{tBi}$, there would be no destructive interference, the cross
section could only grow with the effective coupling, and the lower
and upper limits would have the same size.  If the interference is
strong enough, the minimum of $\sigma/\sigma_{\rm SM}$ can lie below
the lower edge of the allowed band, $1 - 2\varepsilon_\sigma$, as
happens for $t\bar t$ production at $\sqs = 3$~TeV.  In that case, as
the Wilson coefficient grows, the cross section first drops below the
band, then rises again and re-enters it, so that two distinct
intervals appear along each axis, and the allowed region is an
elliptical annulus (Fig.~\ref{fig:contours_r}).
At $\sqs = 6$, $10$, and $14$~TeV, the
best limits come from $t\bar t h$ production.  Its amplitude also
interferes strongly with the SM, but its cross-section uncertainty
is so large that the minimum of $\sigma/\sigma_{\rm SM}$ stays above
$1 - 2\varepsilon_\sigma$, and the allowed region is a solid ellipse
(see Table~\ref{tab:incertezas}).

As an additional observation, $t\bar t h$ production is more
sensitive than $t\bar t$ production at higher energies, despite its
much larger cross-section uncertainty.  This is because of a chiral
suppression of order $m_t^2/s$ in $t\bar t$, which is lifted in
$t\bar t h$ by the additional chirality flip from the Yukawa
coupling.

\section{Final remarks}
\label{sec:conclusions}

\newcounter{aux}

\renewcommand{\theaux}{(\arabic{aux})}

We have estimated individual limits on the Wilson coefficients of
the dimension-six operators of the Warsaw basis~\cite{grz10} that
generate top-quark--electroweak-boson couplings at a muon collider.
Of the nine fermion-boson coefficients, we focus on six; for
$C^{(+)}_{\vphi Q}$ (LEP and SLC), $C_{tG}$ (LHC), and $C_{bW}$
($b\bar b$ and $b\bar b h$ production), top production at the MuC is
not competitive with existing limits.  We have
considered the most important tree-level production modes with top
quarks in the final state, at center-of-mass energies of $3$, $6$,
$10$, and $14$~TeV, with integrated luminosities
${\cal L} = E^2_{\rm cm}/10~\ab1$.  The limits were estimated from
$2\sigma$ cross-section uncertainties, corresponding to $95\%$ CL,
for muon rapidity ranges $|y| < 2.5$, $4$, and $6$ in the
processes with neutral-gauge-boson fusion diagrams.

The electroweak dipole coefficients $C_{tB}$ and $C_{tW}$ are the
couplings to which we find the strongest sensitivity, see
Table~\ref{tab:ctwctbfromtt} in Sect.~\ref{sec:limtt} and
Table~\ref{tab:ctwctbfromtth} in Sect.~\ref{sec:limtth}, where we also
provide a comparison to the limits in the previous literature. For
these dipole coefficients we have also provided two-dimensional
allowed regions and marginalized limits in
Sect.~\ref{sec:marginalized}. Limits on the coefficients
$C^{(-)}_{\vphi Q}$, $C_{\vphi t}$, $C_{t \vphi}$ and $C_{\vphi tb}$
are provided in Tables \ref{tab:limttnunu}--\ref{tab:limsingleth} in
Sect.\ \ref{sec:limttnn}--\ref{sec:limsingleth}.
Furthermore, the individual limits on those six coefficients, for
both the real and imaginary parts of the complex ones, are collected
in Appendix~\ref{app:limits}, where we give the five most sensitive
processes in each case, together with the effect of an assumed
systematic uncertainty.
The limits on $C_{bW}$ from top production are omitted there
because they are weaker by factors of about $5$ to $10$ than those
from $b\bar b$ and $b\bar b h$ production in the literature, as shown
in Table~\ref{tab:limsinglet} and discussed in
Sect.~\ref{sec:limsinglet}.

Our main findings are summarized in the following list.
\begin{trivlist}
  \refstepcounter{aux}
\item[\theaux\label{a1}] For the Yukawa coefficient $C_{t \vphi}$, the
  most sensitive process at $\sqs \geq 6$~TeV is
  $t\bar t \nu \bar \nu$ production, in agreement with
  previous studies \cite{miranda22,wangyukawa}; at $3$~TeV,
  $t\bar t h$ production is more sensitive for the real part.

  \refstepcounter{aux}
\item[\theaux\label{a2}] The $s$-channel processes $t\bar t$ and
  $t\bar t h$ provide the most stringent bounds on the dipole
  couplings $C_{tB}$ and $C_{tW}$: $t\bar t$ production at
  $\sqs = 3$~TeV for the real parts, and $t\bar t h$ production
  otherwise.  This is in agreement with the conclusion of
  Ref.~\cite{han24}. In addition, single-top and single-top-plus-Higgs
  production provide bounds on the dipole couplings that are more
  stringent than the current LHC limits at $\sqs \geq 6$~TeV, and for
  $C_{bW}$ at all energies. Compared with the current CMS
  intervals in Table~\ref{tab:cmsatlaslim}, the allowed intervals for
  these dipole couplings are narrower by factors of about $3$ to $6$
  at $3$ and $6$~TeV, and by an order of magnitude or more at $10$ and
  $14$~TeV.

  \refstepcounter{aux}
\item[\theaux\label{a3}] The left- and right-handed $ttZ$ couplings
  $C^{(-)}_{\vphi Q}$ and $C_{\vphi t}$ obtain their best bounds from
  $t\bar t\nu\bar\nu$ production.  These bounds are narrower than the
  current LHC limits by factors of about $4$ to $5$ at $3$~TeV, and by
  more than an order of magnitude from $6$~TeV on, provided that the
  forward muons are tagged, so that the $t\bar t\mu^+\mu^-$ events can
  be separated.  If the muons are not tagged, the sensitivity to
  $C^{(-)}_{\vphi Q}$ is essentially unchanged, but the allowed
  interval for $C_{\vphi t}$ is essentially no narrower than the LHC
  one at $3$~TeV and narrower by a factor of about $2.5$ at $6$~TeV,
  growing to more than an order of magnitude at $10$ and $14$~TeV.
  
  \refstepcounter{aux}
\item[\theaux\label{a4}] The process
  $\mu^+ \mu^- \to t\bar t \mu^+ \mu^-$ is not very sensitive to any
  of the couplings we have considered.  However, when the outgoing
  muons escape detection ($|y(\mu^\pm)| > 2.5$), this process adds
  to $t\bar t \nu \bar \nu$ production, and the sensitivity
  of the combined process is somewhat degraded compared to that of
  $t\bar t \nu \bar \nu$ alone (see item \ref{a3}).

  \refstepcounter{aux}
\item[\theaux\label{a5}] The right-handed $tbW$ coupling
  $C_{\vphi tb}$ obtains its best limits from single-top and
  single-top-plus-Higgs production. These limits are narrower than the
  current LHC limits by a factor of about $1.5$ at $3$~TeV, growing to
  more than an order of magnitude at $14$~TeV.

\end{trivlist}

Finally, we remark that this analysis provides an overview of the
different production processes and the potential sensitivity to the
various couplings, at the level of undecayed top quarks.  More
detailed analyses, including decayed top quarks, backgrounds, and
kinematic distributions, and in some cases also parton showering,
hadronization, and fast detector simulation, have been made for the
Yukawa coupling, but not for the other five coefficients treated in
Appendix~\ref{app:limits}.
Moreover, the limits presented here are based on statistical
uncertainties only; an assumed systematic uncertainty, as shown in
Appendix~\ref{app:limits}, can weaken considerably the limits from
processes with large cross sections. More realistic analyses will be
required in future studies.

\paragraph*{Acknowledgments}

We acknowledge support from Sistema Nacional de Investigadores de
Secihti, M\'exico.
The authors are grateful for the computing time granted by Lancad and
Secihti on the Super-Computing Hybrid Cluster ``Xiuhc\'oatl'' at the
General Coordination Of Information And Communication Technologies
(CGSTIC) at Cinvestav (http://clusterhibrido.cinvestav.mx).
We acknowledge the use of Anthropic Claude for assistance with
syntax and grammar.

\appendix

\section{Individual limits on Wilson coefficients}
\label{app:limits}

\begin{figure}[t]
  \centering
  \includegraphics[scale=0.5]{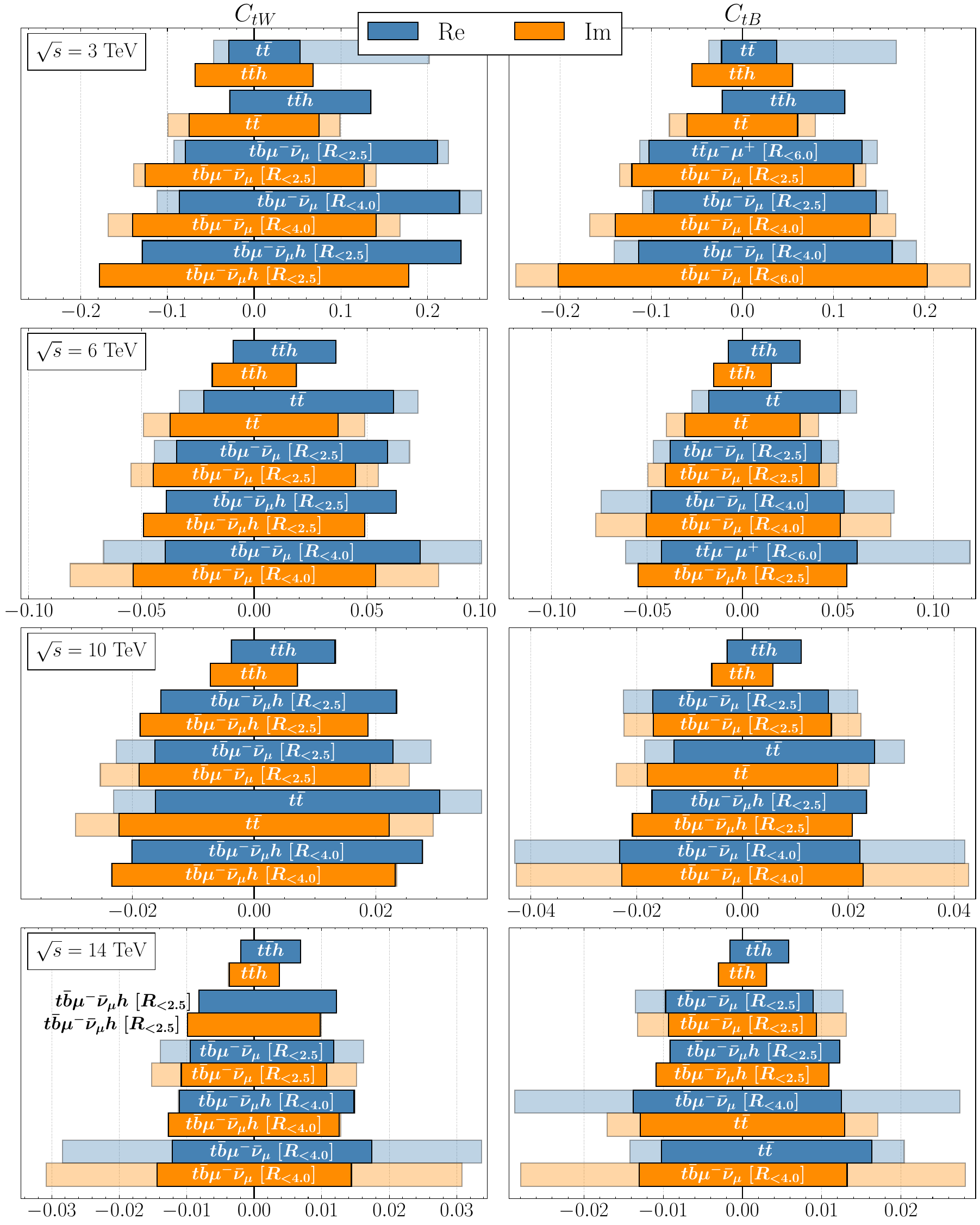}
  \caption{Individual limits on the real and imaginary parts of the dipole
couplings $C_{tW}$ and $C_{tB}$. Solid central bars represent limits
from the cross-section statistical uncertainties
(Table~\ref{tab:incertezas}), while semi-transparent extensions
include an assumed 2\% systematic uncertainty added in quadrature.}
  \label{fig:limits_1}
\end{figure}

\begin{figure}[t]
  \centering
  \includegraphics[scale=0.5]{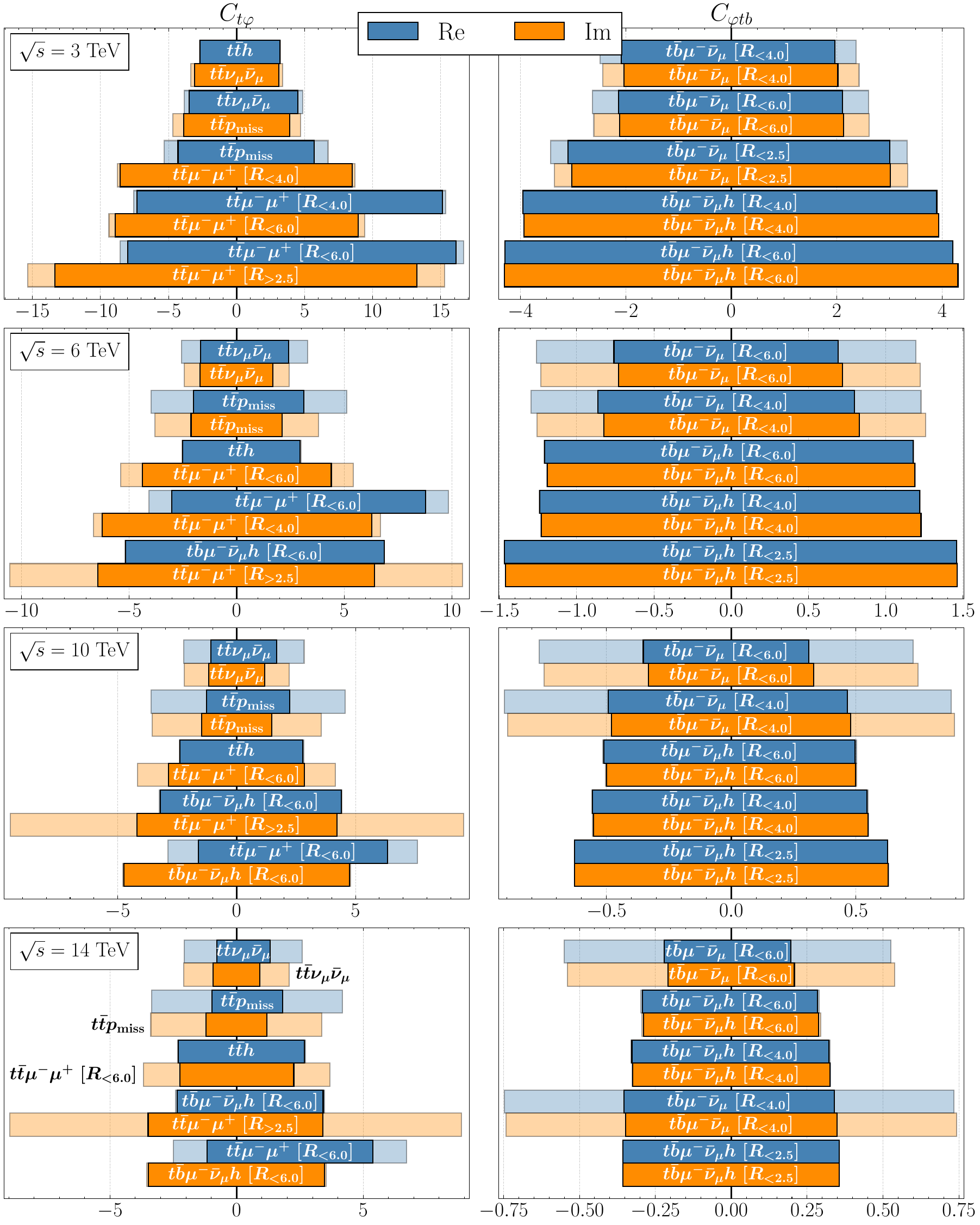}
  \caption{Individual limits on the real and imaginary parts of the
    Wilson coefficients $C_{t\varphi}$ and $C_{\varphi tb}$. The label
    ``$t\bar t\nu_\mu\bar\nu_\mu$'' refers to the process of
    Eq.~(\ref{eq:top.pair.nunu}); the contributions of $\nu_e$ and
    $\nu_\tau$ are negligible.  The label ``$t\bar t p_{\rm miss}$''
    refers to the process
    $\mu^+\mu^- \to t\bar t\mu^+\mu^- + t\bar t\nu\bar\nu$ with
    $|y(\mu)| > 2.5$.  Color code as in Fig.~\ref{fig:limits_1}.}
  \label{fig:limits_2}
\end{figure}

\begin{figure}[t]
  \centering
  \includegraphics[scale=0.5]{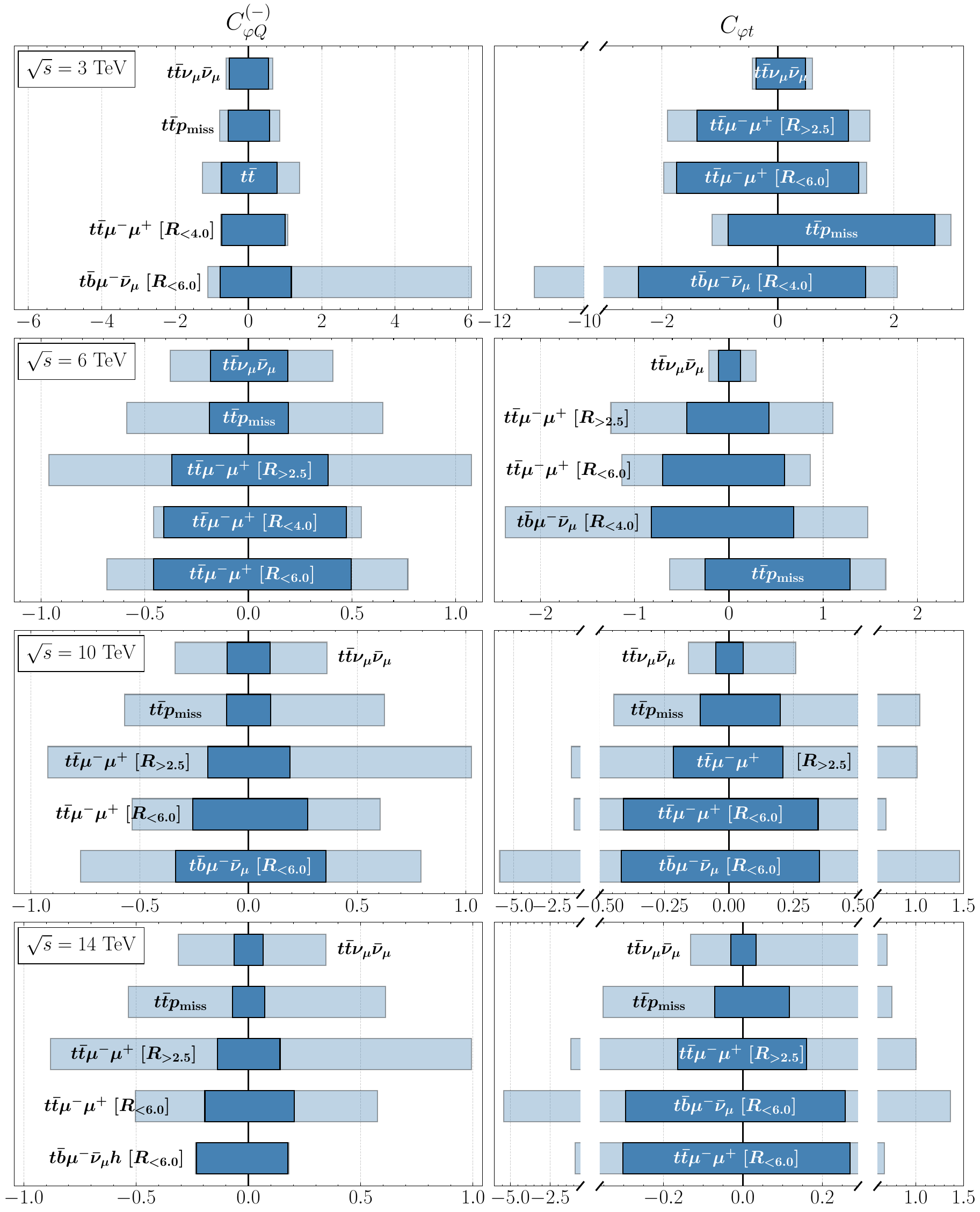}
\caption{Individual limits on the Wilson
coefficients $C_{\varphi Q}^{(-)}$ and $C_{\varphi t}$. Color code as in
Fig.~\ref{fig:limits_1}. The labels ``$t\bar t\nu_\mu\bar\nu_\mu$''
and  ``$t\bar{t}p_\mathrm{miss}$'' as defined in
Fig.~\ref{fig:limits_2}.} 
  \label{fig:limits_3}
\end{figure}

This Appendix presents the individual bounds on the Wilson
coefficients $C_{tW}$, $C_{tB}$, $C_{t\varphi}$, $C_{\varphi tb}$,
$C_{\varphi Q}^{(-)}$, and $C_{\varphi t}$, obtained in this work. The
bounds are shown in Figs.~\ref{fig:limits_1}, \ref{fig:limits_2},
and~\ref{fig:limits_3}.  For each coefficient and center-of-mass
energy $\sqs$, the figures show the five processes that provide the
strongest individual limits with the uncertainties given in
Table~\ref{tab:incertezas}. The limits on $C_{bW}$ obtained from
single-top production are much weaker than those from other processes,
as reported in Sect.~\ref{sec:limsinglet} above, and are therefore not
included here.

Notice that the uncertainties in Table~\ref{tab:incertezas} are
obtained from (\ref{eq:stat.signif}) and are therefore statistical
uncertainties.  This is consistent with the practice in the references
cited in Tables~\ref{tab:hllhc}, \ref{tab:ctwctbfromtt},
and~\ref{tab:ctwctbfromtth}, and allows us to compare our results with
those in these references. For completeness, the figures in this
Appendix also present limits obtained with an assumed systematic
uncertainty of $2\%$ ($1\sigma$), added in quadrature to the
statistical uncertainty $\varepsilon_\sigma$ (half of the values in
Table~\ref{tab:incertezas}); the total is then multiplied by $2$ to
obtain the $2\sigma$ uncertainty used for the limits.  These bounds
are represented by semi-transparent bar extensions.  As discussed in
Sect.~\ref{sec:limsingleth}, the effect is largest for single-top
production at high energies, where the statistical uncertainties are
smallest, and small for single-top-plus-Higgs production, which is
statistically dominated.

The limits given here for systematic uncertainties of $0\%$ and $2\%$
serve as benchmarks that can be used to linearly extrapolate to higher
systematic uncertainties.  This is relevant because at present there
is no realistic estimate of the systematic uncertainties of a
muon-collider detector. We have explicitly checked that, in the
cases where those two limits are substantially different, linear
extrapolation to a systematic uncertainty of $5\%$ gives accurate
results.

\end{document}